\pdfoutput=1

\documentclass[a4paper,11pt]{article}
\usepackage{jheppub}
\usepackage[T1]{fontenc} 
\usepackage{graphicx}
\usepackage{epsfig}
\usepackage{rotating}
\usepackage{amssymb}
\usepackage{dsfont}
\usepackage{psfrag}
\usepackage{amsmath,euscript,array,mathrsfs}
\usepackage{bbold}
\usepackage{epsf}
\usepackage{slashed}
\usepackage{cancel}

\usepackage{caption,subcaption,wrapfig}
\usepackage[colorlinks=true,linktocpage=true,linkcolor=blue,citecolor=blue]{hyperref}

\usepackage{tikz}
\usetikzlibrary{decorations.markings,decorations.pathmorphing}

\newcommand{\be}{\begin{equation}}
\newcommand{\ee}{\end{equation}}
\newcommand{\beqs}{\begin{eqnarray}}
\newcommand{\eeqs}{\end{eqnarray}}

\newcommand{\dd}{\mathrm{d}}

\newcommand{\FF}{\mathcal{F}}
\newcommand{\OO}{\mathcal{O}}
\newcommand{\GG}{\mathcal{G}}

\newcommand{\NN}{\mathcal{N}}
\newcommand{\LL}{\mathcal{L}}
\newcommand{\BB}{\mathcal{B}}
\newcommand{\B}{\mathbb{B}}

\newcommand{\R}{\mathbb{R}}
\newcommand{\CP}{\mathbb{C}\mathbb{P}}
\newcommand{\parent}[1]{\left(#1\right)}
\newcommand{\rhot}{\overline{\rho}}
\newcommand{\Bconf}{\mathbb{B}_8^{\rm{conf}}}

\newcommand{\ffunc}{\mathsf{F}}
\newcommand{\ls}{\ell_s}

\makeatletter
\def\@hex@@Hex#1%
{\if a#1A\else \if b#1B\else \if c#1C\else \if d#1D\else
	\if e#1E\else \if f#1F\else #1\fi\fi\fi\fi\fi\fi \@hex@Hex}
\makeatother

\definecolor{c1}{HTML}{bd4008}
\definecolor{c2}{HTML}{f78026}
\definecolor{c3}{HTML}{fdbe11}
\definecolor{c4}{HTML}{44c721}
\definecolor{c5}{HTML}{1cad85}
\definecolor{c6}{HTML}{2792b6}
\definecolor{c7}{HTML}{025394}
\definecolor{c8}{HTML}{1d3585}
\definecolor{c9}{rgb}{.5,0,.5}

\title{Is entanglement a probe of confinement?}

\author[1,2]{Niko Jokela,}
\author[3]{Javier G. Subils}

\affiliation[1]{Department of Physics, P.O. Box 64, FI-00014 University of Helsinki, Finland}
\affiliation[2]{Helsinki Institute of Physics P.O. Box 64, FI-00014 University of Helsinki, Finland}
\affiliation[3]{Departament de F\'isica Qu\`antica i Astrof\'isica \& Institut de Ci\`encies del Cosmos (ICC), \\ Universitat de Barcelona, Mart\'i Franqu\`es 1, ES-08028, Barcelona, Spain.}

\emailAdd{niko.jokela@helsinki.fi}
\emailAdd{jgsubils@fqa.ub.edu}

\date{\today}

\abstract{We study various entanglement measures in a one-parameter family of three-dimensional, strongly coupled Yang-Mills-Chern-Simons field theories by means of their dual supergravity descriptions. A generic field theory in this family possesses a mass gap but does not have a linear quark-antiquark potential. For the two limiting values of the parameter, the theories flow either to a fixed point or to a confining vacuum in the infrared. We show that entanglement measures are unable to discriminate confining theories from non-confining ones with a mass gap. This lends support on the idea that the phase transition of entanglement entropy at large-$N$ can be caused just by the presence of a sizable scale in a theory.
and just by itself should not be taken as a signal of confinement. We also examine flows passing close to a fixed point at intermediate energy scales and find that the holographic entanglement entropy, the mutual information, and the $F$-functions for strips and disks quantitatively match the conformal values for a range of energies.} 

\preprint{$\begin{array}{rr}
	\text{HIP-2020-29}\\\text{ICCUB-20-023}\end{array}$
}

\begin{document}

\maketitle
\flushbottom

\section{Introduction}\label{sec:intro}
\setcounter{page}{2}

Entanglement entropy is a non-local quantity that permits us to study non-perturbative phenomena of quantum field theories. Despite of its simple definition, it turns out to be very difficult to compute in interacting field theories. The situation is also conceptually challenging in gauge field theories due to the lack of local tensor product decomposition of the physical Hilbert space of gauge invariant states ${\mathcal{H}}_\text{phys}\bcancel{\rightarrow} {\mathcal{H}}_A\otimes {\mathcal{H}}_B$ \cite{Ghosh:2015iwa}. In holography, the Ryu--Takayanagi (RT) formula \cite{Ryu:2006bv} is conjectured to provide us with the entanglement entropy for a given field theory at strong coupling and in the limit of large-$N$. The RT formula states that the entanglement entropy associated with a region $A$ is given by the minimal area of a co-dimension two bulk surface exploring the dual 10-dimensional classical background geometry, anchored onto $\partial A$ at the boundary of the bulk spacetime.

Our main interest in this paper is the question of whether the holographic entanglement entropy as given by the RT prescription can reveal if a gauge field theory is confining. This question was raised by several works which found evidence for phase transitions in entanglement entropies as functions of relevant length scales in different confining backgrounds at large-$N$ \cite{Klebanov:2007ws,Nishioka:2006gr,Nishioka:2009un,Kol:2014nqa}. The phase transitions additionally share some resemblance with the deconfinement phase transitions happening in the same models at finite temperature. This pronounces the expectation that entanglement entropy probes confinement. Interestingly, a number of studies have scrutinized the AdS/CFT results by a comparison to entanglement entropies calculated from 4d pure glue non-Abelian Yang-Mills theories on a lattice with a small number of colors \cite{Buividovich:2008gq,Buividovich:2008kq,Itou:2015cyu,Rabenstein:2018bri}.

In the paper at hand we wish to challenge this paradigm. To do so, we will make clear what we mean by confining behavior. We will say that a theory is confining 
if it exhibits linearly growing potential between quarks and anti-quarks at large separation. As is well-known, there are also other notions of confinement which do not need to agree with this. For example, one could insist on the spectrum only containing color singlets \cite{Greensite:2017ajx}. It is important to keep in mind that in the presence of flavor degrees of freedom there is no local order parameter for confinement since these phases can be continuously connected with Coulomb and Higgs phases (see, however, recent work \cite{Cherman:2020hbe}). At large-$N$, however, one could consider the pressure $p$ to play the role of the order parameter for deconfinement \cite{Thorn:1980iv,Witten:1998zw}: $p\sim{\mathcal{O}}(1)$ for confining phase while 
$p\sim{\mathcal{O}}(N^{2})$ in the deconfining phase in the theories that we consider in this paper. Indeed, the abrupt ``vanishing'' of some quantity counting the number of degrees of freedom has been widely considered a hallmark of confinement in holographic literature. We will revisit the question whether the vanishing of the (regularized) holographic entanglement entropy should be considered as an imprint of the underlying deconfinement-confinement phase transition.

Because the quark-antiquark potential is given by an appropriately chosen Wilson loop, we are essentially going to discuss how both holographic quantities, namely the entanglement entropy and the Wilson loop, should be viewed on a different footing. Let us first recall the more familiar case at finite temperature. From the holographic computation we learn that whenever there is a horizon, the RT surfaces have maximal extensions from the boundary towards the interior of bulk spacetime. On the other hand, at sufficiently high temperature, hanging strings associated with various Wilson loops do not probe the deep interior of the bulk spacetime, since they tend to break apart reflecting thermal creation of quark-antiquark pairs in the gauge theory side. 
The RT surfaces, on the other hand, cannot similarly break apart due to homology constraint and they keep creeping towards the black hole horizon at large distances, eventually matching with the Bekenstein-Hawking entropy density per unit area. This is on par with the common lore that boundary measurements are insensitive to acausal events beyond the entanglement wedge \cite{Headrick:2014cta}.\footnote{Recent attempts at probing beyond the entanglement wedge include explicit construction of judicious Wilson loop configurations \cite{Bao:2019hwq} and entanglement islands in the context of black hole evaporation in JT gravity (see \cite{Almheiri:2020cfm} for a nice review).} Then, holographic entanglement entropy seems to be the ideal tool for bulk reconstruction from boundary measurements of lattice simulations of strongly coupled gauge theories \cite{Jokela:2020auu}, since the RT surfaces have maximal extensions from the boundary towards the interior of bulk spacetime.

Keeping in mind that the geometric realizations of the Wilson loops and entanglement entropies behave rather differently at large temperature we wish to examine what can be learned at the opposite limit of small temperature. We assume that the temperature will be much smaller than any other energy scale of the system and will specifically be interested in gauge field theories which are not scale free. We will probe these gauge theories with Wilson loops and entanglement measures and ask if the linear quark-antiquark potential aka confinement is captured uniquely by the entanglement entropy and, more importantly, whether the analysis of the entanglement entropy dictates if the underlying gauge theory is confining.

We will consider a one-parameter family of three-dimensional gauge theories via their gravity duals. Even though these gauge theories do not explicitly follow from D-brane constructions in Type IIA language, we understand their most important features relevant to the present study. The interactions of the theories are Yang-Mills like, associated to a two-sites quiver, similarly to the Klebanov-Witten (KW) theory \cite{Klebanov:1998hh}. Both gauge groups are in addition supplemented with Chern-Simons (CS) terms. The parameter distinguishing the members of this family, denoted by $b_0\in[0,1]$, is related to the inverse squared couplings of each of the two gauge groups. The gravity duals preserve $\NN=1$ supersymmetry and can be described both in Type IIA and M-theory frameworks, though only the latter enjoys a regular prescription. Based on earlier work by \cite{Cvetic:2001ye,Cvetic:2001pga}, these gravity duals were recently studied in \cite{Faedo:2017fbv}.

The family of solutions as parametrized by $b_0$ allows us to smoothly deform the IR properties. As a result, and depicted in Fig.~\ref{fig:triangle}, the single SYM-CS theory at the UV follows different trajectories along the renormalization group (RG)  down to a phase with a mass gap (except for $b_0=0$) with the possibility of also being confining at the limiting case $b_0=1$. The qualitatively different behaviors for the quark-antiquark potentials for $0<b_0<1$ versus $b_0=1$ is later shown in Fig.~\ref{fig.WilsonLoop}, with only the latter case showing linear dependence on the separation of the pair. This dissimilarity was attributed in \cite{Faedo:2017fbv} to the fact that when $b_0=1$ the CS interactions disappear. When $b_0\neq 1$, however, CS interactions are turned on and it is a natural expectation \cite{Dunne:1998qy} that they give a mass to the gauge bosons, in which case the color charge is consequently screened \cite{Karabali:1999ef}.

Previous works have emphasized that the phase transition of the entanglement entropy at large-$N$ signals the deconfinement transition of the underlying gauge theory and
can be used as a diagnostic for gauge theories with varying degree of resemblance to QCD, see, \emph{e.g.}, \cite{Lewkowycz:2012mw,Kol:2014nqa,Mahapatra:2019uql,Knaute:2017lll,Arefeva:2020uec}. However, 
sometimes the mere presence of an energy scale leads to a phase transition of the entanglement entropy \cite{Klebanov:2012yf,Jokela:2019tsb} in phases with no link to confinement. One of the main results of this work is to demonstrate by explicit examples that various entanglement entropy measures, while being sensitive to the presence of a mass gap, do not distinguish confining from non-confining theories contrary to the quark-antiquark potential. This means that the mass gap fixes the maximum typical distance between entangled states, which a priori is independent from the interactions between infinitely massive and very distant quarks. None of the entanglement measures we investigate is able to distinguish between the confining theory from those that are only gapped. Similarly, we can assert that entanglement measures are not sensitive to the presence of CS interactions. On the contrary, we show that the near-proximity of the conformal fixed point at some intermediate energy scale is clearly captured, in addition to by the entanglement entropy itself, by the mutual information and the $F$-functions counting the numbers of degrees of freedom in $(2+1)$-dimensional field theories.

The rest of the paper is organized as follows. First, in Section~\ref{sec:background} we review the properties of the gravitational solutions \cite{Faedo:2017fbv} and outline salient details of their gauge theory duals. In Section~\ref{sec:EE} we apply standard holographic techniques to compute entanglement entropies for strips and in disks. Having computed these quantities for several representatives of the family, we show how the two limiting $b_0=0,1$ behaviors are attained in Section~\ref{sec:limiting}. We finally discuss the results and conclude in Section~\ref{sec:conclusions}. Many technicalities are relegated to appendices.

\section{Background solution}\label{sec:background}

Our starting point is the one-parameter family of supersymmetric gravity solutions constructed in \cite{Faedo:2017fbv}, whose salient features we review in this Section. Along the way, we will also discuss their gauge theory duals.
We choose these background geometries as our arena since they provide us with explicit realizations of gravity duals of theories possessing phases that are gapped but not necessarily confining.  We believe that our message below can be adapted more generally, however.
 
The solutions of \cite{Faedo:2017fbv} are induced by a stack of $N$ coincident M2-branes with a transverse space which is one of the eight-dimensional manifolds belonging to the $\B_8$ class, found originally in \cite{Cvetic:2001ye,Cvetic:2001pga}. They have Spin(7) holonomy so that they preserve $\NN = 1$ supersymmetry and form a one-parameter family of solutions, characterized by a scale arising from the fact that there is a four-cycle whose size remains finite while the geometry smoothly caps off. The size of this four-cycle maps to a non-vanishing mass gap in the dual field theory.

In ``cigar-like'' geometries such as here, the natural expectation is that the dual field theory is in a confining phase in the sense of that an external quark-antiquark pair feels a linear attractive potential at large separation \cite{CasalderreySolana:2011us}. This expectation stems from the fact that in the holographic computation of the potential between the quark pair \cite{Maldacena:1998im,Rey:1998ik}, in which one considers a string attached  
to them at the boundary of spacetime, the potential is essentially given by the length of the string: the geometry ending smoothly causes the string to find it advantageous to dive steeply and stretch in the spatial direction only at the bottom of the geometry, once the quark and the antiquark are sufficiently separated. A crucial ingredient in this way of reasoning is that a string cannot break into two disconnected pieces due to charge conservation. This is because the string endpoints carry charges and therefore they need to end on branes. The string cannot end on thin air. This holographic description hence provides us with a nice geometric picture of the notion that a quark or an antiquark cannot exist in isolation in a confining phase.

The gauge theories studied in \cite{Faedo:2017fbv} provide a counterexample to such an expectation, as we shall explain next. The key reason is that the regular low-energy supergravity description is available in eleven-dimensional M-theory but not in ten-dimensional Type IIA theory. The technical reason behind this is that in eleven dimensions isolated quarks are described by membranes wrapping the M-theory circle, that would from string picture correspond to endpoints carrying charges. Membranes have boundaries. If the M-theory circle caps off smoothly at the end-of-space, the membranes wrapping this cycle can loose one of their boundaries in a smooth manner. In such a situation, issues that would otherwise be related to charge conservation are avoided and ``separated'' pairs of quarks are allowed. This phenomenon leads to a theory with a mass gap but with no obvious signal of confining behavior. From the field theory point of view, what seems to be allowing the quarks to be in isolation is the presence of CS interactions. It is well known that in three dimensions CS interactions induce a mass for the gauge bosons. This causes the screening of color charges and allows the fluxtube between quarks to break. Let us next review the construction of the geometry and then explain how these phenomena are geometrically realized in these models.

\begin{figure}[t]
	\centering
	\begin{tikzpicture}[scale=3.3,very thick,decoration={markings,mark=at position .5 with {\arrow{stealth}}}]
	\node[above] at (0,0) {SYM-CSM $|$ D2};
	\node[below] at (0,-2.2) {Mass gap $|$ $\mathbb{R}^4\times {\rm S}^4$};
	\node[left,red] at (-1,-1) {OP $|$ CFT};
	\node[right] at (2,-2) {Confinement $|$ $\mathbb{R}^3\times{\rm S}^1\times {\rm S}^4$};
	\draw[postaction={decorate},ultra thick,c4] (0,0) --  (0,-2) node[left,midway]{$\mathbb{B}_8$};
	\draw[postaction={decorate},ultra thick, gray] (-1,-1) -- (-1,-2) node[left,midway]{$\mathbb{B}_8^{\textrm{\tiny OP}}$\,\,};
	\draw[postaction={decorate},ultra thick, c2] (0,0) .. controls (-.9,-.9) and (-1,-1) .. (-0.98,-2);
	\draw[postaction={decorate},ultra thick, c3] (0,0) .. controls (-.5,-.5) and (-.5,-1.5) .. (-.5,-2);
	\draw[postaction={decorate},ultra thick, c5] (0,0) .. controls (.5,-.5) and (.5,-1.5) .. (.5,-2);
	\draw[postaction={decorate},ultra thick, c6] (0,0) .. controls (.9,-.9) and (.95,-1.05) .. (1,-2);
	\draw[postaction={decorate},ultra thick, c7] (0,0) .. controls (.9,-.9) and (1.5,-1.6) .. (1.5,-2);
	\draw[postaction={decorate},ultra thick, c8] (0,0) .. controls (.95,-.95) and (1.8,-1.8) .. (1.9,-2);
	\draw[|-|] (-1,-2) -- (0,-2);
	\draw[-stealth] (0,-2) -- (2,-2);
	\node[left=5] at (-1,-2) {$b_0$};
	\node[below=5] at (-1,-2) {$0$};
	\node[below=5] at (0,-2) {$2/5$};
	\node[below=5] at (2,-2) {$1$};
	\draw[postaction={decorate},ultra thick,c1] (0,0) -- (-1,-1) node[left,midway]{$\mathbb{B}_8^\infty$\,};
	\draw[postaction={decorate},ultra thick,c9] (0,0) -- (2,-2) node[right,midway]{\, $\mathbb{B}_8^{\rm{conf}}$};
	\draw [red, ultra thick,fill=red] (-1,-1) circle [radius=0.03]; 
	\draw [black, fill=black, ultra thick] (0,0) circle [radius=0.03];
	\end{tikzpicture}
	\caption{\small Pictorial representation of the $\B_8$ family of solutions. The asymptotic UV regime is given by the 3D super Yang-Mills theory with Chern-Simons interactions for the gauge fields (SYM-CSM). As we come down in energy (descending in the plot) the RG flow generically drives the theory to an IR regime with a mass gap. Only for extreme value of $b_0=1$ do we flow to a confining theory, as depicted on the bottom-right corner of the plot. For $b_0=0$ the IR is governed by an Ooguri-Park (OP) conformal fixed point. The hue or the warmth of the curves will be roughly in one-to-one correspondence with the values of $b_0$ on the horizontal axis; in the figures to follow we will try to maintain this mapping.
	}\label{fig:triangle}
\end{figure}
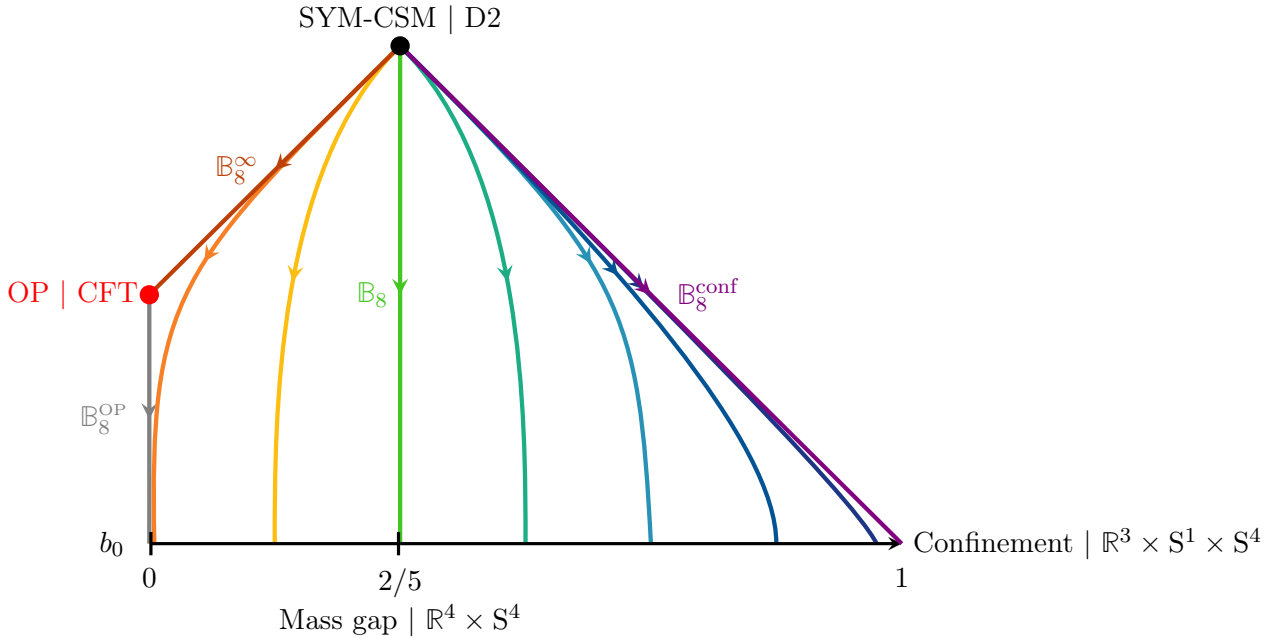

\begin{figure}[t]
	\begin{center}
		\begin{subfigure}{0.45\textwidth}
			\includegraphics[width=1.\textwidth]{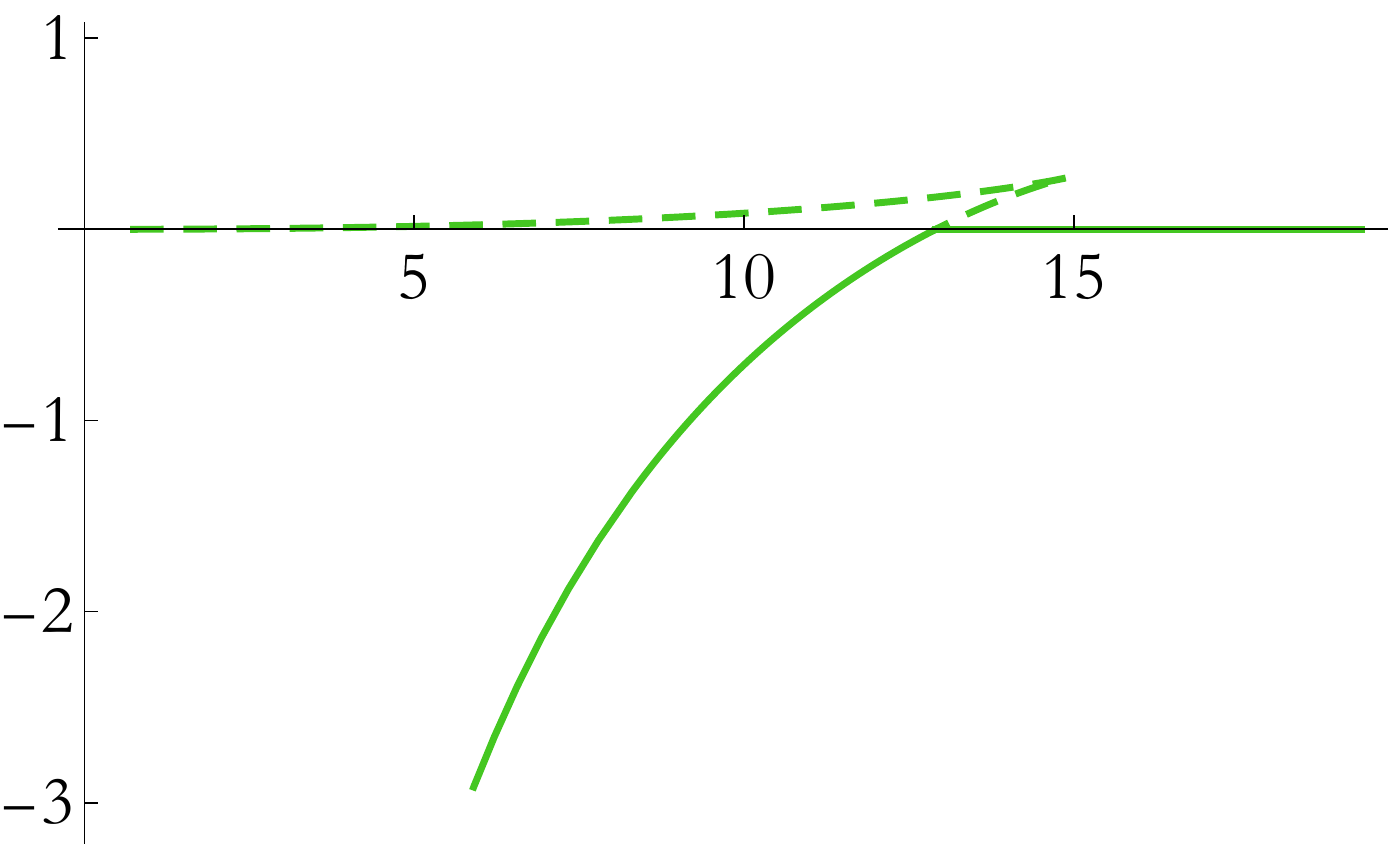} 
			\put(-200,125){$V/M_0$}
			\put(-35,65){$10^2 d/L_0$}
		\end{subfigure}\hfill
		\begin{subfigure}{.45\textwidth}
			\includegraphics[width=1.\textwidth]{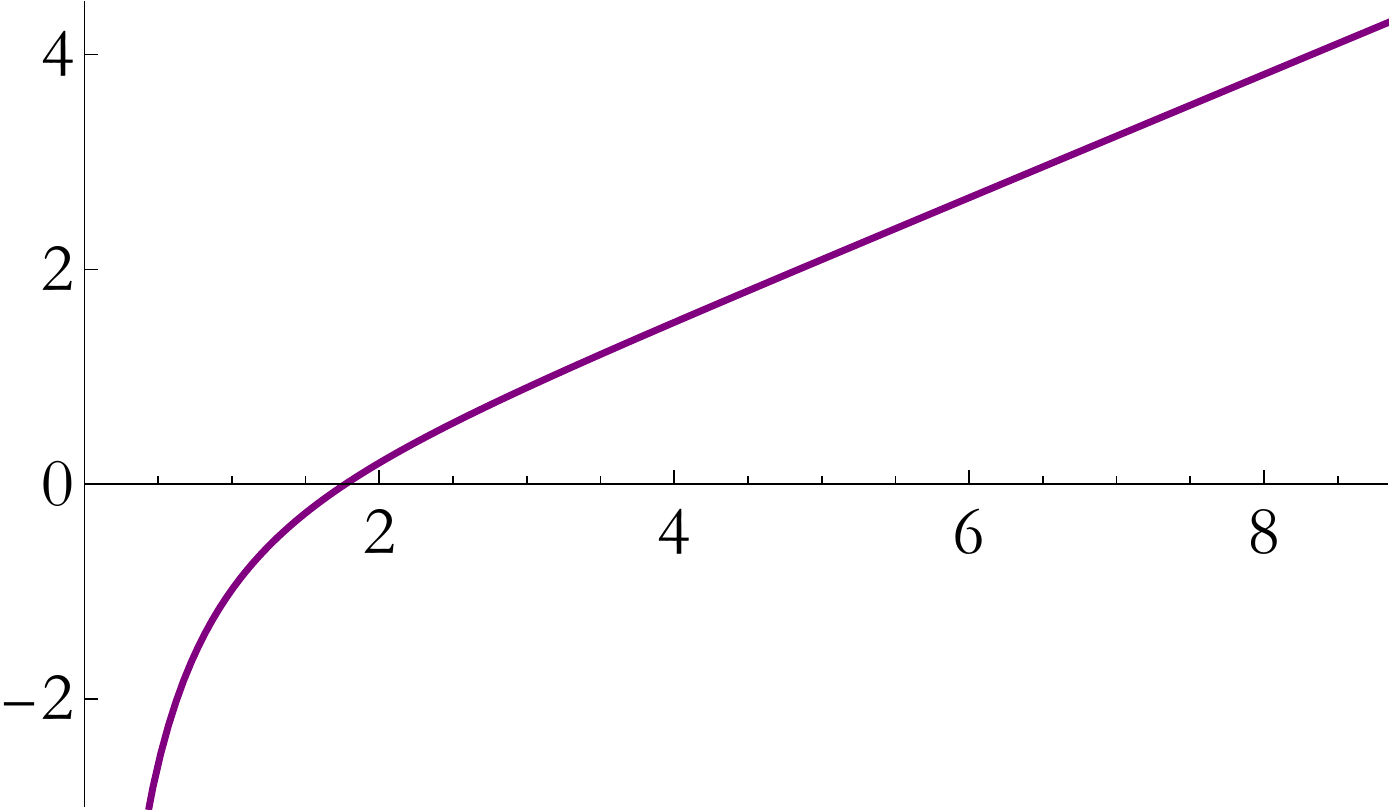} 
			\put(-200,125){$V/M_0$}
			\put(-25,20){$d/L_0$}
		\end{subfigure}
		\caption{\small Quark-antiquark potential for a non-confining theory with a mass gap $\B_8^0$ (Left) and for the confining theory $\Bconf$ (Right). Solid curves stand for the value of the potential for the dominant configuration whereas dashed curves depict those of unstable configurations. The string breaks in the $\B_8^0$ theory when the curve crosses zero, signaling the splitting of a meson into two quarks in the gauge theory. As explained in the main text, splitting cannot happen in the confining case $\Bconf$, where consequently the connected configuration is always the dominant one, leading to the linear growth of the potential for large values of the separation between the quark and the antiquark. Plot adapted from \cite{Faedo:2017fbv} and in the units used therein.}\label{fig.WilsonLoop}
	\end{center}
\end{figure}

Bearing in mind that the regular description is only provided within 11d supergravity, we are nevertheless going to work in type IIA supergravity since the UV of these solutions is better understood in ten dimensions. The string-frame metric takes the form:
\begin{eqnarray}\label{10Dansatz}
 \dd s_{\rm st}^2 &=&h^{-\frac12}\, \dd x_{1,2}^2 +h^{\frac12}\bigg(\dd r^2+e^{2f}\dd\Omega_4^2+e^{2g}\left[\left(E^1\right)^2+\left(E^2\right)^2\right] \bigg) \ ,
\end{eqnarray}
together with a non-trivial dilaton $e^\Phi = h^{\frac{1}{4}}e^\Lambda$. In \eqref{10Dansatz}, $r$ is the holographic radial coordinate, $E^1$ and $E^2$ describe a two-sphere ${\rm S}^2$ fibration over the four-sphere ${\rm S}^4$. This four cycle is the key player in our analysis and corresponds to the one that does not pinch off in the IR of the M-theory realization, the volume form of which is $\dd\Omega_4^2$ (see Appendix~\ref{ap:solution} for more details). In this Ansatz, the UV of the entire family of geometries is nothing but that induced by the stack of $N$ coincident D2-branes in the decoupling limit:
\begin{equation}
e^{2f}\,=\,2\,e^{2g}\,\sim\, r^2\qquad , \qquad e^{\Phi}\,\sim\,h^\frac14\qquad , \qquad h\,\sim\,N\,r^{-5}\ .
\end{equation}
Notice that since $e^f\neq e^g$, the internal manifold is described by the squashed Fubini-Study metric of $\CP^3$. Notice also that the squashing is radially dependent with asymptotic boundary value $e^{2f-2g} = 2$ for all the solutions, corresponding to the nearly K\"ahler point of $\CP^3$. As proposed in \cite{Loewy:2002hu}, the gauge theory dual of it consists of a two-sites Yang-Mills quiver with $\rm U(N)\times \rm U(N)$ gauge group and bifundamental matter, reminiscent to the KW quiver in four dimensions \cite{Klebanov:1998hh}. In the system at hand, the non-trivial fibration of the circle on which we reduce gives us a non-vanishing internal two-form, generating CS interactions in the gauge theory dual. On top of that, additional internal three- and four-form fluxes in our setup signal the presence of fractional D2-branes, which we expect to produce a shift $M$ in the rank of one of the gauge groups \cite{Aharony:2008gk}. Consequently, these gravity solutions are describing RG flows in a 
\begin{equation}\label{quiver}
{\rm U}(N)_k\times{\rm U}(N+M)_{-k}
\end{equation}
quiver gauge theory with CS interactions at level $k$ while preserving $\NN = 1$ supersymmetry. 

Finally, the one-parameter family of solutions is labeled by a parameter denoted by $b_0$. This is the asymptotic UV value of the NS two-form flux on a two-cycle within $\CP^3$. It has a direct interpretation in the gauge theory side, controlling the difference between the microscopic Yang-Mills couplings of each of the two factors in the gauge group
\begin{equation}\label{eq:bparameter}
 b_0\,\sim\,\frac{1}{g_1^2}-\frac{1}{g_2^2}\ .
\end{equation}
Following the conventions in \cite{Faedo:2017fbv}, this quantity takes values in $b_0\in[0,1]$ and allows us to pictorially represent the family of solutions using $b_0$ as the horizontal axes, as in Fig.~\ref{fig:triangle}.

For a generic value of this parameter, the geometries end smoothly at a certain value $r_s$ of the radial coordinate. This leads to gapped behavior in the corresponding gauge theories due to inherent mass scale in the system.
As depicted in Fig.~\ref{fig.WilsonLoop} (Left), for a generic $b_0$ the quark-antiquark potential does not show linear growth for large separation, watering down signals for confinement.  In fact, from the IR expansion of the metric \eqref{eq:IRexpansionmetric} it is easy to check that the string tension vanishes at $r_s$, despite the absence of a horizon in the geometry. This is the technical reason for the quark pair to split apart.
 
One could argue though that the limiting values 0 or 1 for $b_0$ are special and should be considered on a different footing, since they lead to qualitatively different IR physics. 
But this is precisely at the heart of the matter. When $b_0=1$ we do actually find a confining theory in the sense of a linear growth of the quark-antiquark potential, as depicted in Fig.~\ref{fig.WilsonLoop} (Right). This theory was originally found in \cite{Cvetic:2001ma}. The geometric understanding stems from the fact that in this case the $\rm S^1$ is trivially fibered over the rest of the geometry and consequently does not contract at the IR. This can be attributed to the vanishing of the CS level in \cite{Faedo:2017fbv}, in accordance with the arguments of \cite{Herzog}. This solution is referred to as $\Bconf$. This also leads to the fact that CS interactions are absent, which means that in this case the color charge is not screened from the field theory perspective. It should be noted that we do not claim that this phase is confining in the strict sense. In addition to mapping out the full spectrum, one should also study 't Hooft loops and other higher-rank Wilson loops. These investigations are beyond the scope of this paper, however.

On the other hand, when the difference between the couplings vanishes for  the opposite limiting value $b_0=0$, the mass gap is lost and the theory flows to an IR fixed point described by the Ooguri-Park CFT \cite{Ooguri:2008dk}, which is a deformation of the ABJM theory \cite{Aharony:2008ug} preserving $\NN = 1$ supersymmetry. In this case, the flow is denoted as $\B_8^\infty$. Interestingly, we can pick arbitrarily small but non-vanishing values of $b_0$ in such a way that the RG flow will pass arbitrarily close to the fixed point before deviating towards the gapped IR. This leads to quasi-conformal dynamics in some range of energies, and we shall investigate in Section \ref{sec:limiting} whether any imprint of walking behavior is captured by the analysis of the entanglement measures.

\section{Entanglement entropy}\label{sec:EE}

In order to address the question of whether the entanglement entropy is sensitive to the fact that a theory is confining, we are going to consider different entangling surfaces in the family of solutions we have been discussing in the previous section. Although in the end we want to consider entanglement entropy of strips and disks, we plan to explain first the general setup and specialize to those cases afterwards. Following \cite{Ryu:2006bv}, the entanglement entropy of a QFT region $A$ bounded by $\partial A$ is given by the area of a minimal surface $\Sigma_A$ anchored on $\partial A$ at the spacetime boundary. This minimal surface $\Sigma_A$ is co-dimension two, {\emph{i.e.}}, an eight-dimensional submanifold embedded in our ten-dimensional type IIA geometry \eqref{10Dansatz}, wrapping completely the internal part of the geometry. For simplicity, we will refer to this embedding as the Ryu-Takayanagi (RT) surface associated to $A$. The holographic entanglement entropy in string frame reads
\begin{equation}\label{eq:EE_formula}
 S_A = \frac{1}{4 G_{10}}\ \int_{\Sigma_A} \dd^8\sigma \ e^{-2\Phi}\sqrt{\det \mathsf{g}} \ ,
\end{equation}
where the $\sigma$'s are coordinates on $\Sigma_A$, the constant $G_{10} = (16\pi)^{-1} (2\pi)^7 g_s \ls^8$ is the ten-dimensional Newton's constant, the function $\Phi$ is the dilaton, and $\mathsf{g}$ is the induced metric on $\Sigma_A$ in string frame:
\begin{equation}\label{eq:embedding}
 \mathsf{g}_{\alpha\beta} = \frac{\partial x^{\mu}}{\partial \sigma^\alpha}\frac{\partial x^{\nu}}{\partial \sigma^\beta} \ g_{\mu\nu} \ .
\end{equation}
Since we require that $\Sigma_A$ wraps the whole internal space, integration over six of the eight $\sigma$'s coordinates gives a factor of $V_6 = 32\pi^3 /3$ in \eqref{eq:EE_formula}, which is the volume of $\CP^3$. The embedding of the surface is determined by the equations of motion for the scalar fields that we will choose to present as follows
\begin{equation}
\label{eq:general_embedding}
t= \text{constant} \quad , \quad x_1 = x_1(\sigma_1,\sigma_2) \quad , \quad x_2 = x_2(\sigma_1,\sigma_2) \quad , \quad r = r(\sigma_1,\sigma_2) \ .
\end{equation}
We then vary \eqref{eq:EE_formula} with respect to the fields and obtain the Euler-Lagrange equations
\begin{equation}
\label{eq:EulerLagrange}
\frac{\partial\LL}{\partial\phi^i} - \partial_\mu\left(\frac{\partial \LL}{\partial(\partial_\mu\phi^i)}\right) = 0 \quad , \quad \mbox{for }\phi^i\in\{x^1,\ x^2,\ r\}\mbox{ and }\mu=\sigma^1, \ \sigma^2 \ ,
\end{equation}
which comprise three second order partial differential equations. These equations are the equations of motions that a surface has to fulfil in order to be extremal. We will stick to the entanglement entropy between strips and disks because in that case the problem simplifies and we just have to solve second order ordinary differential equations. Our expectation is that these two cases capture the main features one might encounter considering other shapes for $A$.

This system is quite similar to that of \cite{Bea:2013jxa}, which we followed to perform our computations. We relegate the most technical details to the appendixes and refer there for further details.

\subsection{Entanglement entropy of the strip}\label{sec:strip}

\begin{figure}[t]
	\begin{center}
		\includegraphics[width=\textwidth]{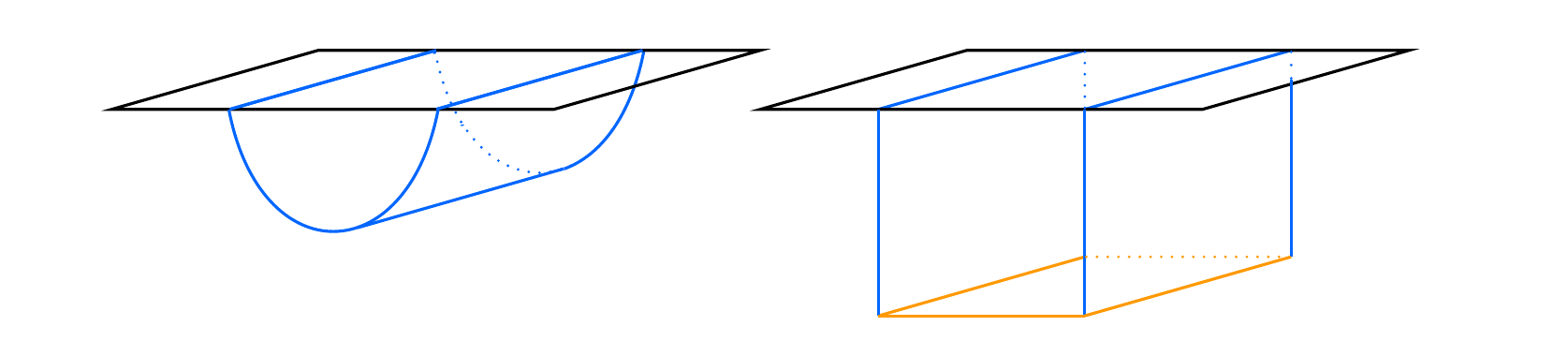} 
			\put(-170,-10){$\sqcup$ configuration}
			\put(-350,-10){$\cup$ configuration}
	\caption{\small  The two possible and competing configurations of the RT surface we have to consider when computing entanglement entropies of strips: for small widths of the strip, a ``connected'' configuration $\cup$, which does not reach the bottom of the geometry (left), competes with the ``disconnected'' configuration $\sqcup$, which reaches the end-of-space (right).
		}\label{fig:strip_conf}
	\end{center}
\end{figure}

Let us now specialize our setup and pick a particular boundary region $A$. Let $A$ be a strip of width $l$. For this choice, we will find two possible embeddings that extremize the area of the RT surface, depicted in Fig.~\ref{fig:strip_conf}. First, there is the configuration denoted by $\cup$, which is specified by the choice
\begin{equation}
\label{eq:embedding_strip1}
t= \text{constant}, \qquad x^1 = \sigma^1\in[-l/2,l/2],\qquad x^2 = \sigma^2\in \R,\qquad r = r(\sigma^1)\in [r_*,\infty) \ ,
\end{equation}
where $r_* \ (> r_s)$ is the value of the radial coordinate at which the RT surface has a turning point. Recall that $r_s$ is the value of the radial coordinate at the end-of-space. When substituting this Ansatz in the equations of motion \eqref{eq:EulerLagrange}, we are left with a unique second order differential equation. Focusing on the induced metric obtained by substituting \eqref{eq:embedding_strip1} into \eqref{10Dansatz}, and taking the squared root of its determinant as specified by \eqref{eq:EE_formula}, we obtain an expression for the entanglement entropy in this case
\begin{equation}
\label{eq:EE_strip}
S_\cup(l) = \frac{V_6 L_y}{4 G_{10}} \int_{-\frac{l}{2}}^{\frac{l}{2}}\dd \sigma^1\  \Xi^{\frac{1}{2}} \left(1+h\ \dot r^2\right)^{\frac{1}{2}}\ ,
\end{equation}
where the dot stands for differentiation with respect to $\sigma^1$, $L_y = \int_{\R}\dd\sigma^2$, and 
\begin{equation}\label{eq:def_XI}
\Xi = h^2 e^{8f+4g-4\Phi}\ .
\end{equation} 
In \eqref{eq:EE_strip} there is a conserved quantity that allows us to solve the remaining second order equation and will help us to perform the integrals involved in the computation of the entanglement entropy. We relegate these details to Appendix~\ref{ap:strip}.

There is also a second possible configuration we denoted by $\sqcup$ in Fig.~\ref{fig:strip_conf}. This extremal surface consists of three pieces: two of them extend from the UV all the way down to the IR, whereas the third one lays at the bottom of the geometry connecting the two. The embeddings for the two former ones are specified by demanding
\begin{equation}
\label{eq:embedding_strip2}
t= \text{constant}\quad , \quad x^1 = \pm \frac{l}{2} = \text{constant}\quad , \quad x^2 = \sigma^2\in \R\quad  , \quad r =\sigma^1 \in [r_s,\infty) \ ,
\end{equation}
which automatically fulfill the equations of motion \eqref{eq:EulerLagrange}. As we pointed out, these embeddings are describing two submanifolds that hang from the UV, where they are attached to one of the edges of the strip; towards the IR, where they end at the regular bottom of the geometry at $r=r_s$. Note that on their own they do not yet constitute a valid RT surface for the strip, since 
they are not homologous to the boundary region $A$ unless they are connected at the bottom with the third piece  mentioned above. This last piece is located at the bottom of the geometry and specified in the following way:
\begin{equation}
\label{eq:bottom}
t= \text{constant}\quad , \quad x^1 = \sigma^1 \in[-l/2,l/2] \quad ,\quad x^2 = \sigma^2 \in \R \quad ,\quad r = r_s = \text{constant}\ .
\end{equation}
It also fulfils equations \eqref{eq:EulerLagrange}. Additionally, it has zero area due to the fact that it is wrapping the four-cycle which  contracts smoothly at the end-of-space. Thus, the action \eqref{eq:EE_formula} in this second RT surface $\sqcup$ consequently leads to the expression 
\begin{equation}
\label{eq:EE_disconnected}
S_\sqcup =  2 \ \frac{V_6 L_y}{4 G_{10}} \int_{r_s}^{\infty}\dd  r\  \Xi^{\frac{1}{2}} h^{\frac{1}{2}}\ 
\end{equation}
for the entanglement entropy. Notice that (\ref{eq:EE_disconnected}) does not {\emph{explicitly}} depend on $l$. The dependence on $l$ is only through the boundary conditions (\ref{eq:embedding_strip2}) by anchoring the surface to the boundary region $A$.  
The configuration of a single dipping surface of the form (\ref{eq:embedding_strip2}), connected to a semi-infinite bottom embedding, would lead to the entanglement entropy of spacetime divided in half. It is immediate that the corresponding entanglement entropy cannot depend on the position of the domain wall; similar argument applies to (\ref{eq:EE_disconnected}).

\begin{figure}[t]
	\begin{center}
		\begin{subfigure}{0.47\textwidth}
			\includegraphics[width=\textwidth]{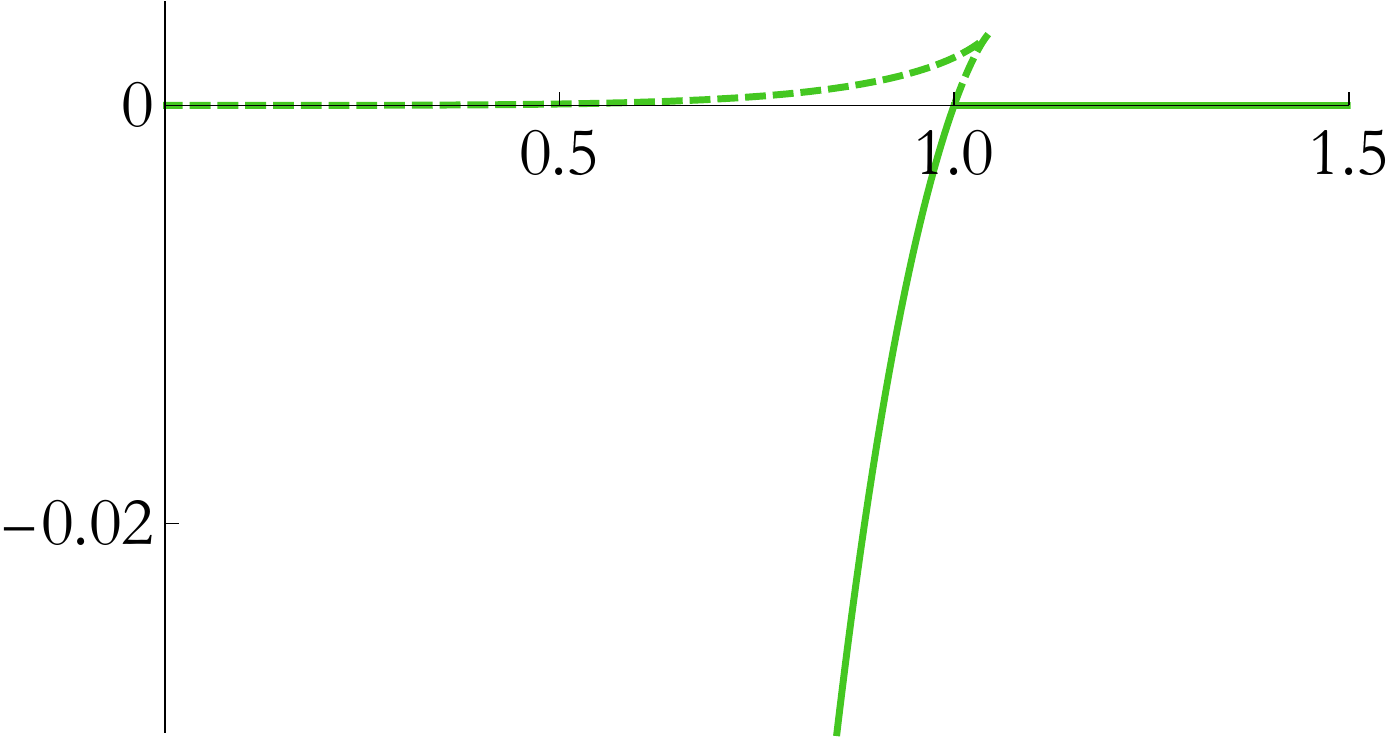} 
			\put(-180,115){$\Delta \overline S$}
			\put(-30,-10){$l/l_c$}
		\end{subfigure}\hfill
		\begin{subfigure}{.47\textwidth}
			\includegraphics[width=\textwidth]{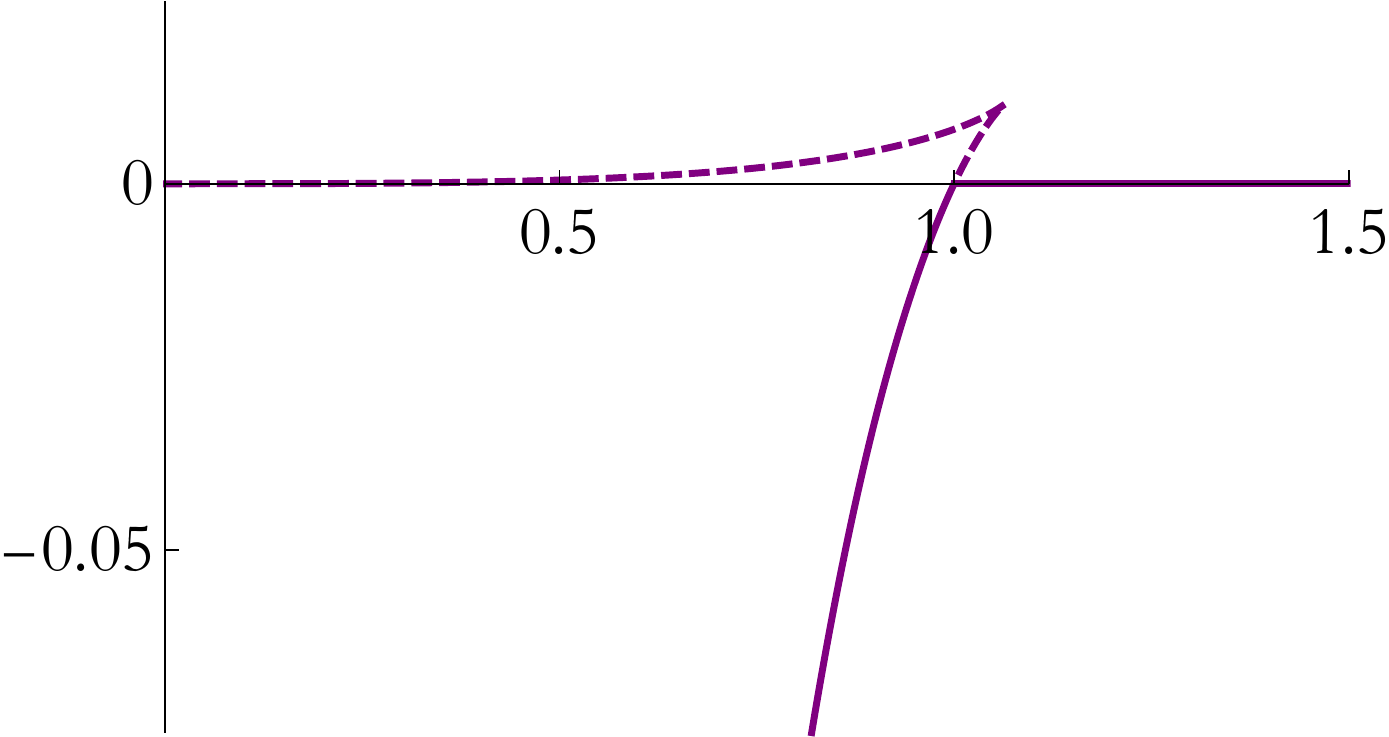} 
			\put(-180,115){$\Delta \overline S$}
			\put(-30,-10){$l/l_c$}
		\end{subfigure}
		\caption{\small Entanglement entropy of a single strip as a function of the width of the strip in the gapped non-confining theory $\B_8^0$ (Left) and in the confining one $\Bconf$ (Right). We plot the rescaled quantity \eqref{eq:dimensionlessEEstrip} defined in Appendix~\ref{ap:strip} as a function of the strip width normalized to the value above which the ``disconnected'' configuration $\sqcup$ becomes the dominant one.
		}\label{fig.EEplotsStrip}
	\end{center}
\end{figure}

Notice that for a given strip of width $l$ we have two candidate RT surfaces (although we will see that the configuration $\cup$ is only present below some critical value of $l$), so we need to infer which one is the minimal one. The correct choice can be expressed as:
\begin{equation}
S_{\text{strip}}(l) = \min\{S_\cup(l),\ S_\sqcup\} \ .
\end{equation}
Notice that both \eqref{eq:EE_strip} and \eqref{eq:EE_disconnected} are UV divergent. However, instead of regulating the entropy functionals for each case, which comes with its own subtleties, we are content with comparing the on-shell actions. To be more precise, we will consider the following object
\begin{equation}
	\label{eq:EE_strip_reg}
	\Delta S (l) = S_\cup (l) - S_{\sqcup}\ .
\end{equation}
Then, while \eqref{eq:EE_strip} and \eqref{eq:EE_disconnected} are UV divergent quantities, since their divergence structure is the same due to homogeneity, \eqref{eq:EE_strip_reg} is a finite quantity. 
Moreover, there will be values for the width of the strip at which $\Delta S (l)$ will flip sign, signalling a critical value above which the ``disconnected'' configuration $\sqcup$ becomes preferred. In fact, we can see that happening in Fig.~\ref{fig.EEplotsStrip}, where we show the value of a dimensionless version of \eqref{eq:EE_strip_reg} defined in Appendix~\ref{ap:strip} as a function of the width of the strip. Above a certain critical value of the width of the strip $l=l_c$ the disconnected configuration becomes the dominant one. On top of that, $\Delta S(l)$ has a turning point causing the disconnected configuration to be the only existing one for large enough widths $l$.

\begin{figure}[t]
	\begin{center}
		\begin{subfigure}{0.45\textwidth}
			\includegraphics[width=\textwidth]{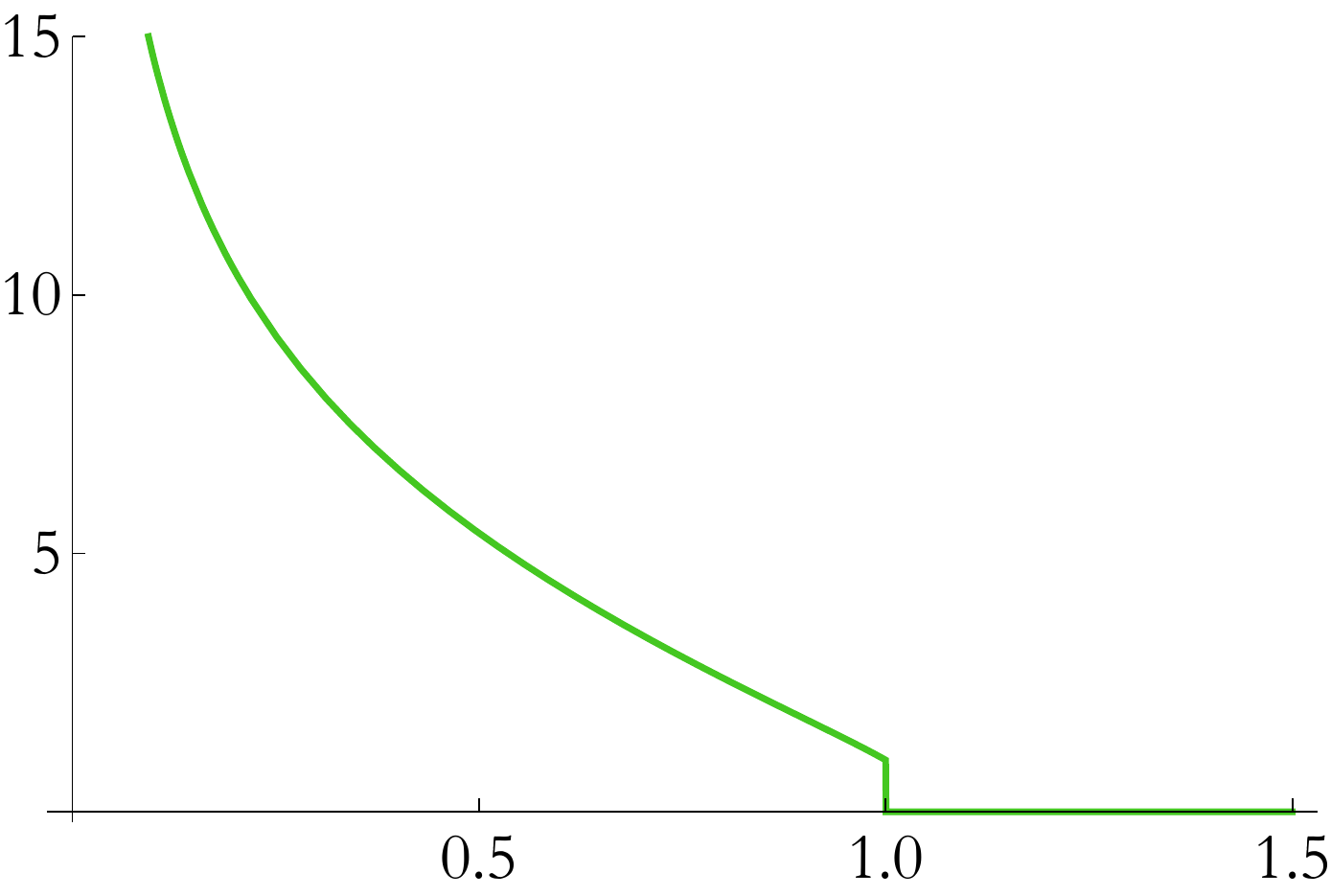} 
			\put(-190,140){$\ffunc(l)/\ffunc(l_c)$}
			\put(-30,-10){$l/l_c$}
		\end{subfigure}\hfill
		\begin{subfigure}{.45\textwidth}
			\includegraphics[width=\textwidth]{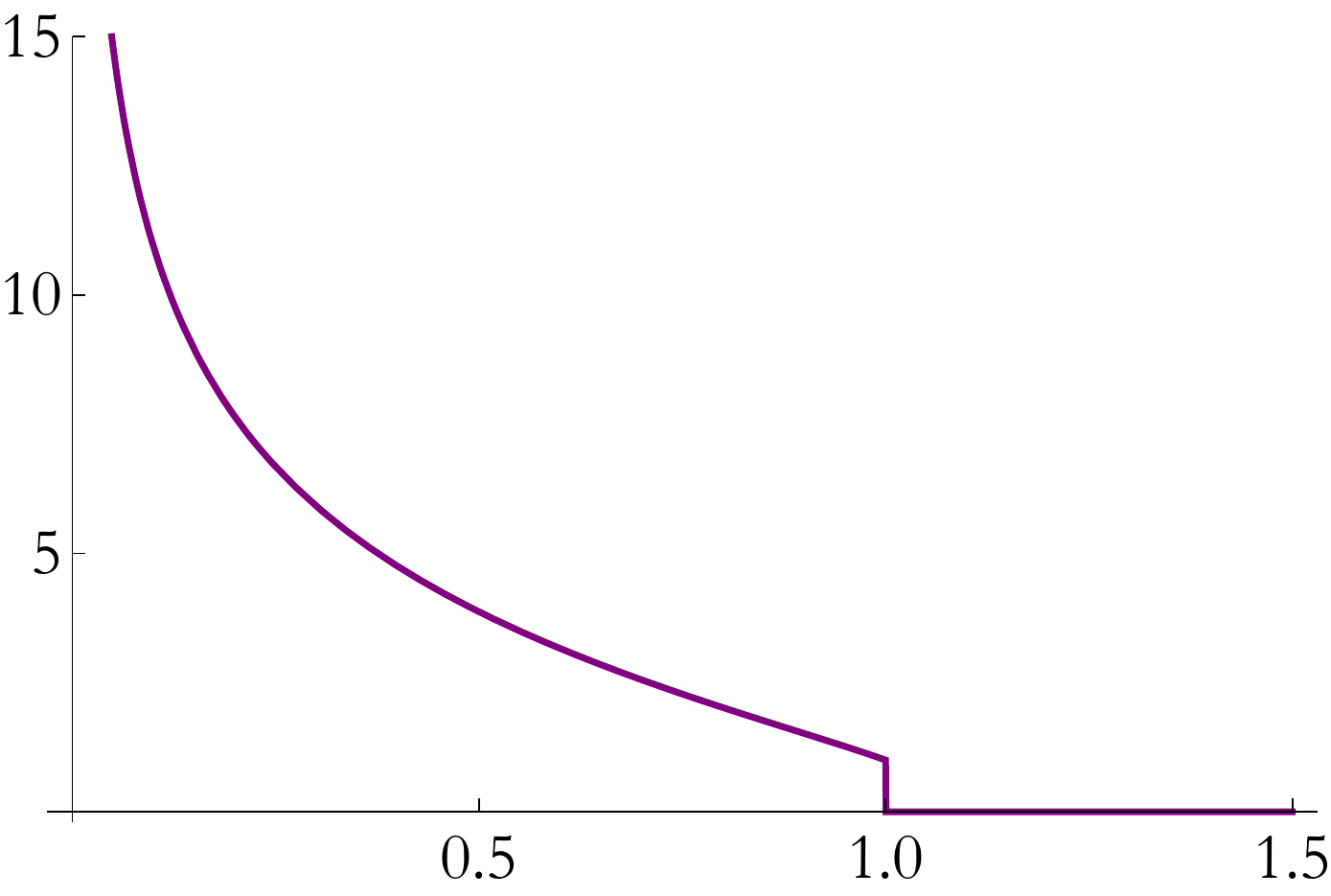} 
			\put(-190,140){$\ffunc(l)/\ffunc(l_c)$}
			\put(-30,-10){$l/l_c$}
		\end{subfigure}
		\caption{\small Function $\ffunc(l)$ for a single strip as a function of strip width in the gapped non-confining theory $\B_8^0$ (Left) and the confining one $\Bconf$ (Right). Both quantities are normalized to their value at the point where the ``disconnected'' configuration $\sqcup$ becomes dominant, above which it is strictly zero.
		}\label{fig.FRplotsStrip}
	\end{center}
\end{figure}

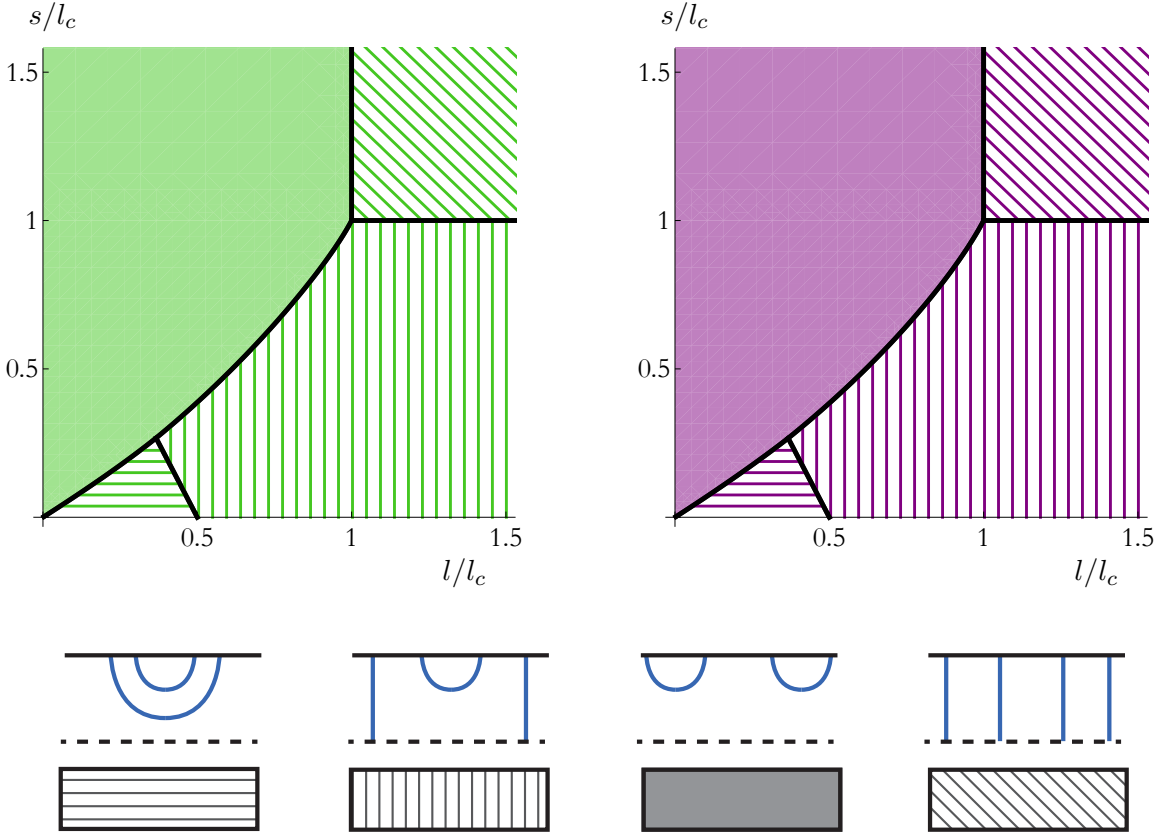
\begin{figure}[t]
	\begin{center}
		\begin{subfigure}{0.45\textwidth}
			\includegraphics[width=\textwidth]{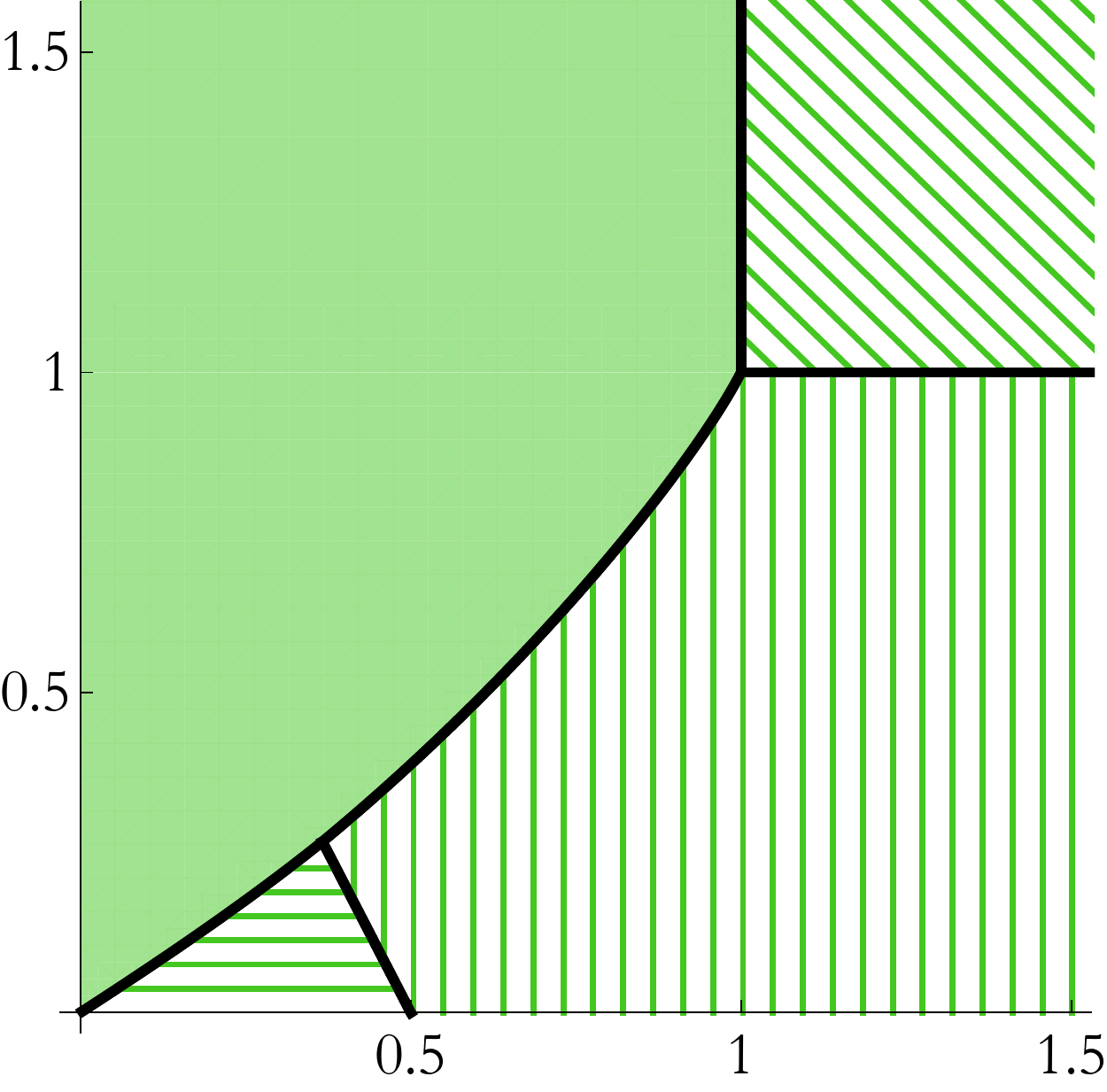} 
			\put(-185,200){$s/l_c$}
			\put(-30,-10){$l/l_c$}
			\put(-30,-35){$ $}
		\end{subfigure}\hfill
		\begin{subfigure}{.45\textwidth}
			\includegraphics[width=\textwidth]{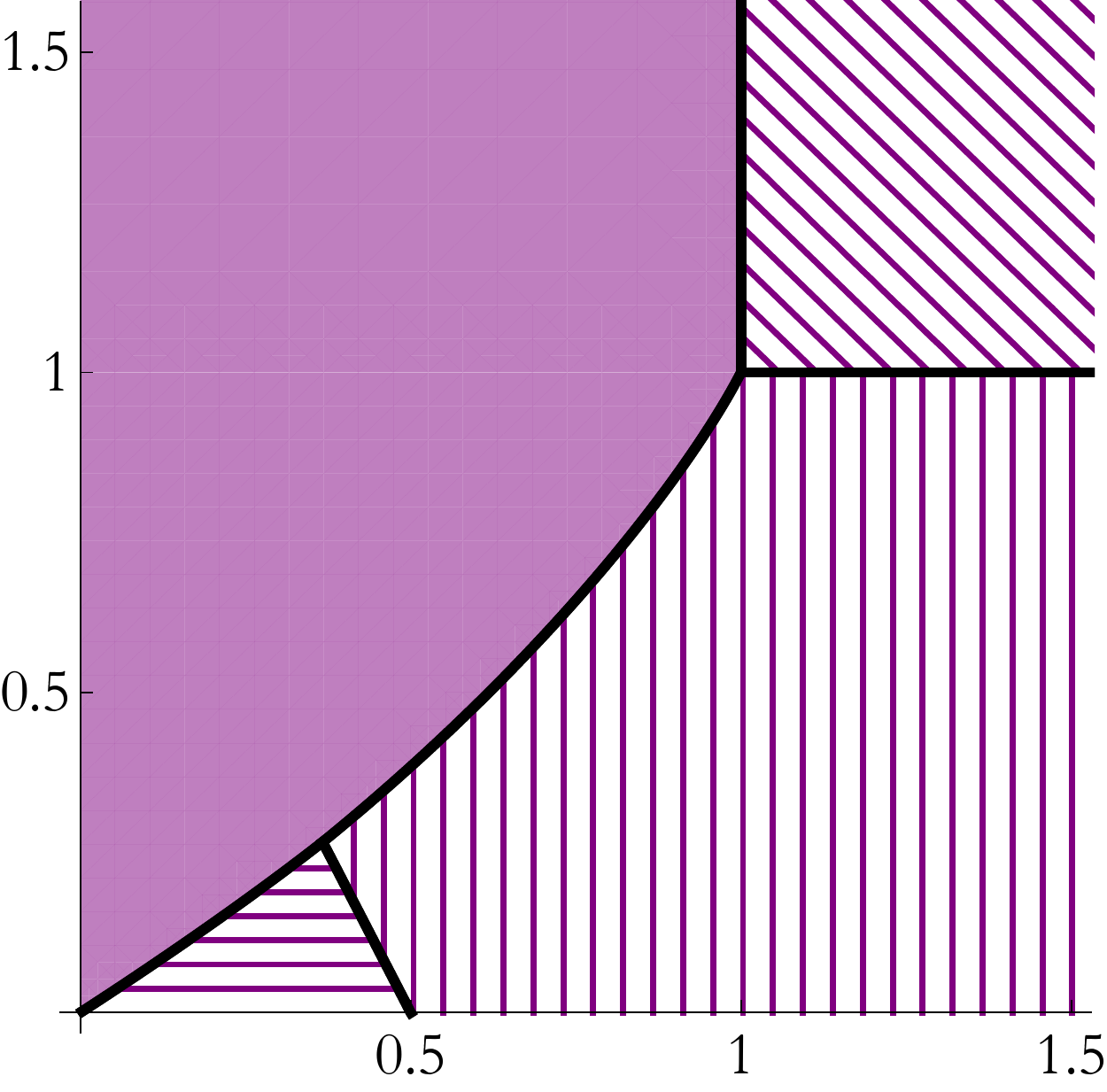} 
			\put(-185,200){$s/l_c$}
			\put(-30,-10){$l/l_c$}
			\put(-30,-35){$ $}
		\end{subfigure}
		\includegraphics[width=\textwidth]{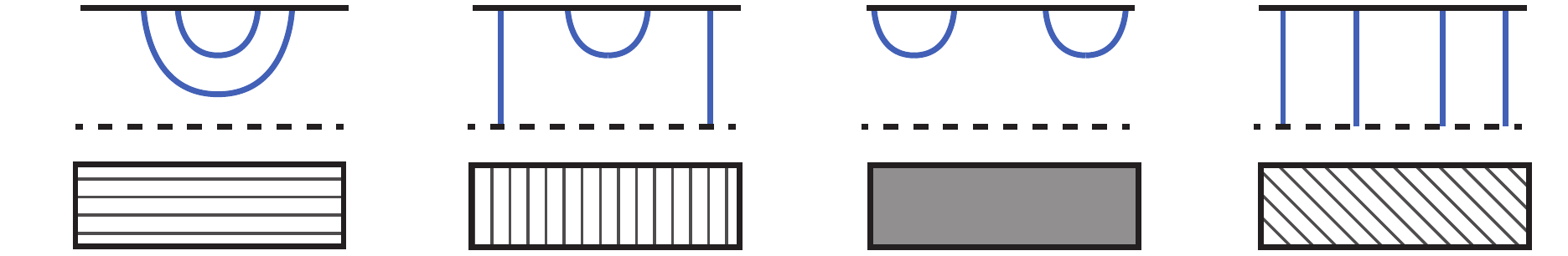} 
		\caption{\small Phase diagram of mutual information of entanglement entropy of strips in the gapped non-confining theory $\B_8^0$ (Left) and the confining one $\Bconf$ (Right). Solid black curves consist of points where mutual information \eqref{eq:mutual_info} vanish and a phase transition takes place. Separation between strips $s$ is shown in the vertical axes and their width $l$ is shown in the horizontal axes, both quantities normalized to the value of the width $l_c$ at which the ``disconnected'' configuration $\sqcup$ becomes dominant in each case.
		}\label{fig.mutualInfostrips}
	\end{center}
\end{figure}

Once the entanglement entropy of a single strip as a function of its width is known, it is straightforward to compute a c-function proposed by Liu and Mezei \cite{Liu:2012eea,Liu:2013una} that is conjectured to measure the number of degrees of freedom at scale  $l$:
\begin{equation}
\label{eq:Ffunction_strip}
\ffunc_{\text{strip}}(l) = \frac{l^2}{L_y}\frac{\partial S_{\text{strip}}}{\partial l} = \frac{l^2}{L_y}\frac{\partial \Delta S}{\partial l}\ ,
\end{equation}
where in the last equality we took into account \eqref{eq:EE_strip_reg} and the fact that $S_\sqcup$ does not explicitly depend on $l$. We plot this quantity in Fig.~\ref{fig.FRplotsStrip}. The behavior of this quantity is an immediate consequence of the behavior of $\Delta S $: it decreases continuously until the point where the connected configuration is disfavored, where it suddenly jumps to zero. For values of the widths bigger than $l_c$, $\ffunc_{\text{strip}}(l)$ is identically zero. 
In this case the RT surface is probing IR scales and is not sensible to massive degrees of freedom which are gapped. 

Let us pause to make a technical comment. Taking numerical derivatives in (\ref{eq:Ffunction_strip}) can be demanding. One can completely circumvent this procedure by using the chain rule explained in \cite{Jokela:2020auu} to find that\footnote{Note the usage of the rescaled dimensionless quantities in favor of $\Delta S$ and $l$ as defined in Appendix~\ref{ap:strip}.}
\begin{equation}
\label{eq:derivative}
\frac{\partial \Delta \overline S}{\partial \overline l} = \Xi_*^{\frac{1}{2}} \ .
\end{equation}
From this expression it is manifest that the computation of $\ffunc_{\text{strip}}$ only involves finite quantities and no UV regularization is invoked. We have explicitly checked that results following from (\ref{eq:derivative}) and (\ref{eq:Ffunction_strip}) agree to great accuracy.

Another interesting quantity to compute is the mutual information between two entangling strips $A$ and $B$, given by 
\begin{equation}\label{eq:mutualpreinfo}
I(A,B) = S_A + S_B - S_{A\cup B} \ .
\end{equation}
The mutual information characterises the amount of information shared by the two domains \cite{Headrick:2010zt}. 
If we consider mutual information between two strips of the same width $l$ which are separated by a distance $s$, the expression (\ref{eq:mutualpreinfo}) can be written as
\begin{equation}
\label{eq:mutual_info}
I(A,\tilde A) = 2 \Delta S(l) - \Delta S (2l+s) - \Delta S (s) \ .
\end{equation}
It is possible then to draw a phase diagram of the dominant phases, the phase boundaries given by the locii where the mutual information vanishes. This is shown in Fig.~\ref{fig.mutualInfostrips}, where it can be seen that four different regions arise, depending on which is the dominant configuration in each case. The result is analogous to those of \cite{Ben-Ami:2014gsa,Jokela:2019ebz} discussing cigar-like geometries. Interestingly, similar four-zone phase diagram is present in anisotropic non-confining geometries \cite{Jokela:2019tsb,Hoyos:2020zeg}, pronouncing the fact that it is the presence of internal mass scale behind the entanglement entropy phase transitions and not necessarily the confinement. 

\subsection{Entanglement entropy of the disk}\label{sec:disk}

\begin{figure}[t]
	\begin{center}
		\includegraphics[width=\textwidth]{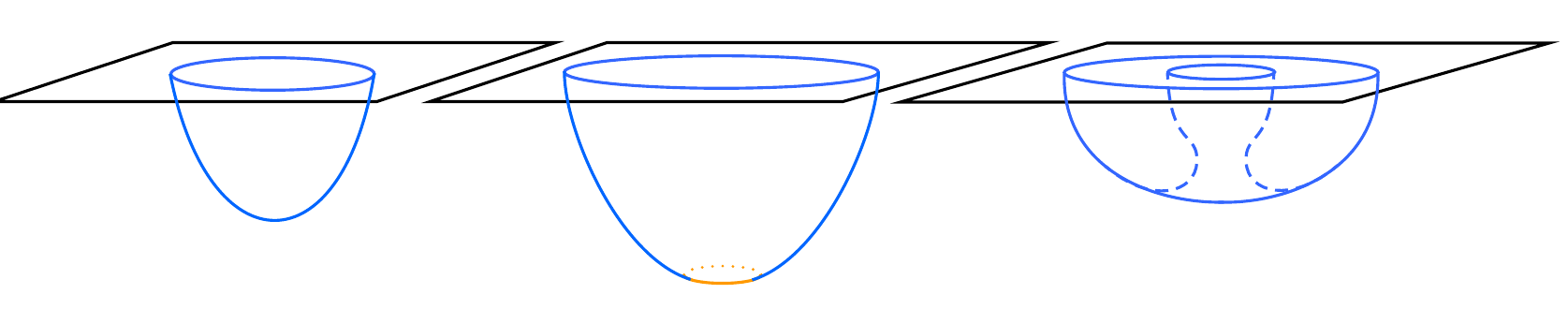} 
			\put(-100,-10){(c)}
			\put(-238,-10){(b)}
			\put(-361,-10){(a)}
		\caption{\small Three types of embeddings found when solving the equation of motion found when computing entanglement entropy on disks. When the radius of the disk is sufficiently small, the embedding does not reach the IR part of the geometry (a). For large values of the radius of the disk, the embedding reaches the end-of-space (b). Additionally, when considering mutual information between disks, embeddings like (c) should be taken into account.
		}\label{fig:disk_conf}
	\end{center}
\end{figure}

Similarly, we can study the case in which the entangling region is a disk of radius $R$ to inspect which lessons from the preceding subsection are intact. In the case of a disk, we perform a change of coordinates to polar coordinates in the gauge theory directions, namely $x^1 = \rho \cos \alpha$, $x^2 = \rho \sin \alpha$. This choice will reduce our system of three second order differential equations obtained from \eqref{eq:EulerLagrange} to a single second order differential equation. Taking that into account, the embedding of the RT surface in our background solutions is going to be determined by the choice
\begin{equation}
\label{eq:embedding_disk}
t= \text{constant} \quad , \quad \rho =\sigma^1 = [0,R] \quad , \quad \alpha = \sigma^2 \in [0,2\pi] \quad , \quad r = r(\sigma^1) \in [r_*,\infty)\ ,
\end{equation}
and such that 
\begin{equation}\label{eq:radiusat infty}
\lim_{\rho\to R} r(\rho) = \infty \ ,
\end{equation} 
which essentially tells us that the embedding is attached to the circumference of the corresponding disk. Satisfying these conditions, there are two distinct types of embeddings, depicted in Fig.~\ref{fig:disk_conf} (a) and (b). One possibility, happening when the radius of the disk is sufficiently small, is that the RT surface does not reach the bottom of the geometry, leading to the condition 
\begin{equation}
\label{eq:disk_BC1}
r(0) = r_* \qquad , \qquad  \dot r(0) = 0 \ .
\end{equation}
This is what happens in Fig.~\ref{fig:disk_conf}(a). Another option is that the embedding does enter all the way down to the IR part of the geometry, in which case the condition is that
\begin{equation}
\label{eq:disk_BC2}
\lim_{\rho\to \rho_*} r(\rho) = r_s \ .
\end{equation}
Recall $r_s$ is the value of the radial coordinate $r$ at the end-of-space. In this case, similarly to the strip configuration, as represented in Fig.~\ref{fig:disk_conf}(b), we need an extra piece of surface laying at the bottom of the geometry in order to complete the RT surface. We already argued above that this piece, specified by \eqref{eq:bottom}, fulfils the equations of motion.

\begin{figure}[t]
	\begin{center}
		\begin{subfigure}{0.45\textwidth}
			\includegraphics[width=\textwidth]{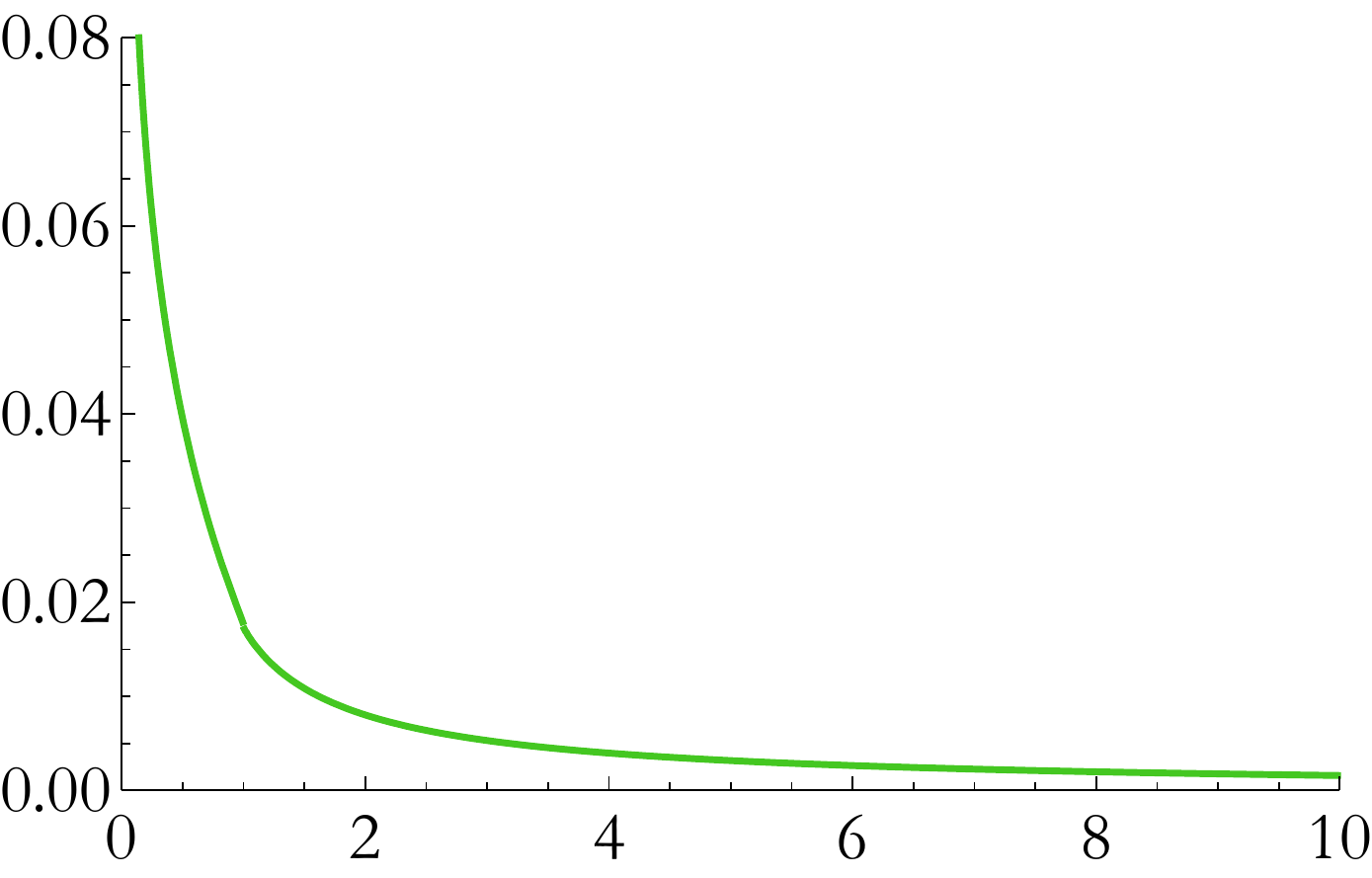} 
			\put(-185,135){$\overline\ffunc_D(R)$}
			\put(-30,-10){$R/R_c$}
		\end{subfigure}\hfill
		\begin{subfigure}{.45\textwidth}
			\includegraphics[width=\textwidth]{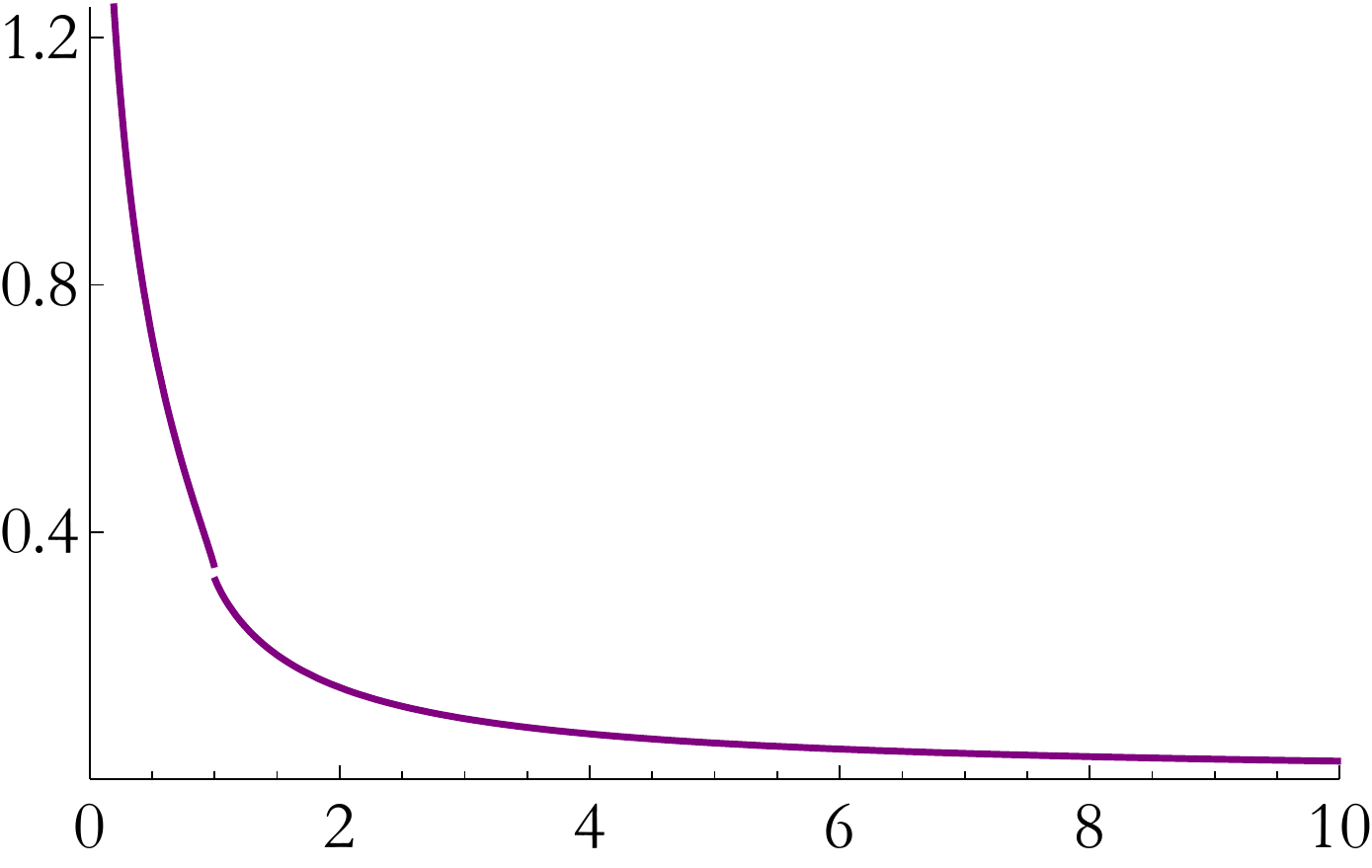} 
			\put(-185,135){$\overline\ffunc_D(R)$}
			\put(-30,-10){$R/R_c$}
		\end{subfigure}
		\caption{\small Measure of the flow of the degrees of freedom in the gapped non-confining theory $\B_8^0$ (Left) and in the confining one $\Bconf$ (Right). Notice that we plot rescaled quantities as explained in Appendix~\ref{ap:disk}, as a function of the radius of the disk normalized to the smallest radius for which the RT surface reaches the end-of-space.
		}\label{fig.FofR_Disk}
	\end{center}
\end{figure}

When substituting \eqref{eq:embedding_disk} into \eqref{eq:EE_formula} we get
\begin{equation}\label{eq:EE_disk}
 S_{\text{disk}}=\frac{V_6}{4G_{10}} \int_0^{2\pi}\dd\sigma^2  \int_0^R \dd \rho \ (1+h\ \dot r ^2)^{\frac{1}{2}}\ \rho\ \Xi^{\frac{1}{2}}
\end{equation}
and the first integral immediately gives us a factor of $2\pi$. The (\ref{eq:EE_disk}) is UV divergent and we will need to regularize it using counterterms, as shown in Appendix \ref{ap:disk}. Having done so, we will refer to the renormalized quantity as $S_{\text{disk}}^{\text{reg}}$. An analogous quantity to \eqref{eq:Ffunction_strip} can be defined for the disk to measure the change in the degrees of freedom \cite{Liu:2012eea}
\begin{equation} \label{eq:Ffunction_disk}
 \ffunc_{\text{disk}}(R) = R\cdot \frac{\dd S^{\text{reg}}_{\text{disk}}}{\dd R} - S^{\text{reg}}_{\text{disk}} \ .
\end{equation}
In Fig.~\ref{fig.FofR_Disk} we plot a rescaled version of this quantity, again for the two theories $\B_8^0$ and $\Bconf$. We cannot rule out the possibility that $\ffunc_{\text{disk}}$ is discontinuous: from our numerical computations it seems there is a tiny jump when the transition between the two configurations happens. This could in principle happen when the embedding starts reaching the IR bottom of the geometry, 
where the altered boundary conditions can lead to a discontinuity in the derivative of the entropy with respect to the radius. A discontinuity would, however, be in conflict with statements in \cite{Klebanov:2012yf}, where this analysis was done in a different albeit similar system to ours and it was claimed that this function has a second-order phase transition. Numerics are quite involved when the transition occurs, so it is difficult for us to distinguish small discontinuities from numerical errors. It would be interesting to generalize the chain rule argument of \cite{Jokela:2019tsb} that was used in the case of strip configurations to bypass numerical artifacts.

\begin{figure}[t]
	\begin{center}
		\begin{subfigure}{0.45\textwidth}
			\includegraphics[width=\textwidth]{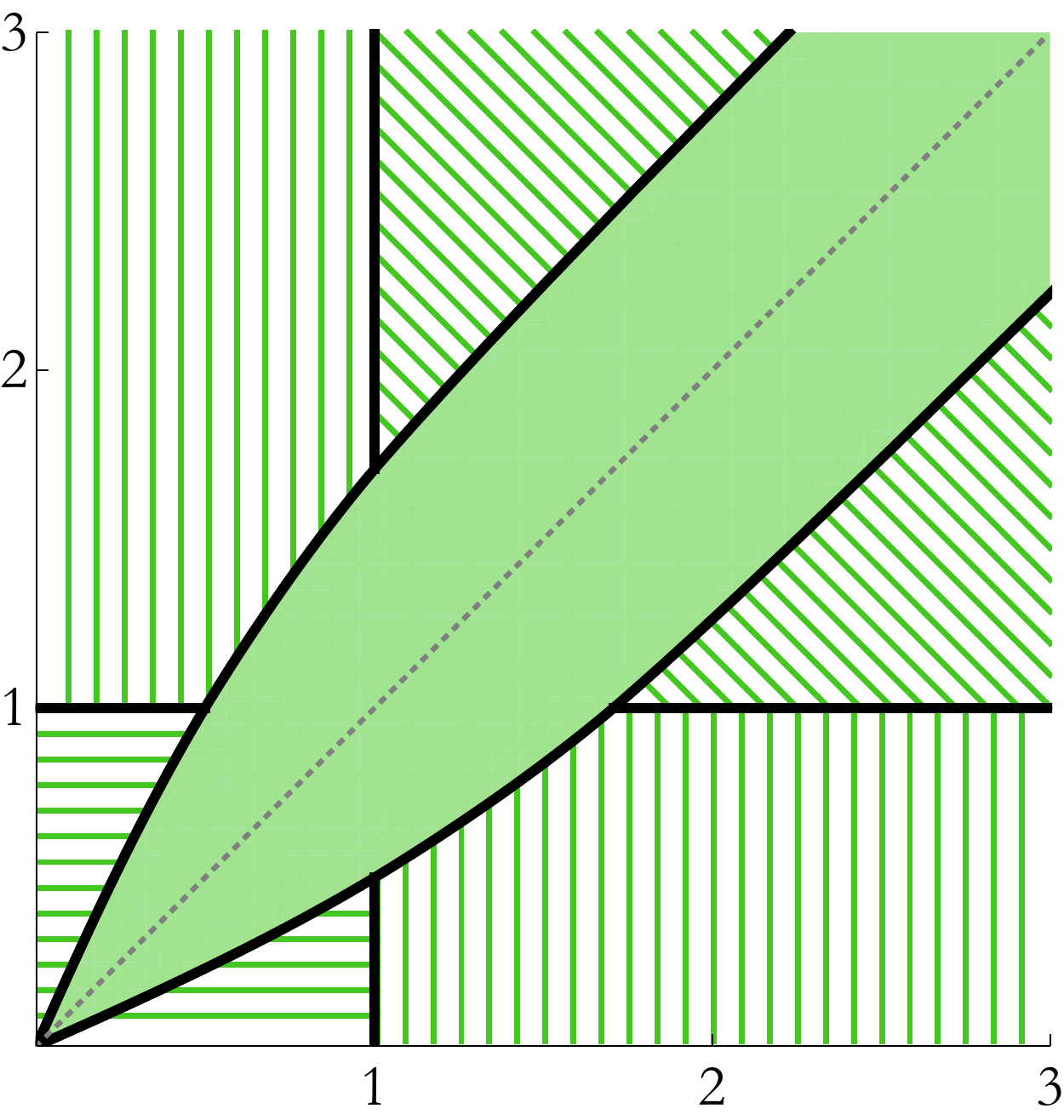} 
			\put(-185,210){$R_2/R_c$}
			\put(-30,-10){$R_1/R_c$}
			\put(-30,-35){$ $}
		\end{subfigure}\hfill
		\begin{subfigure}{.45\textwidth}
			\includegraphics[width=\textwidth]{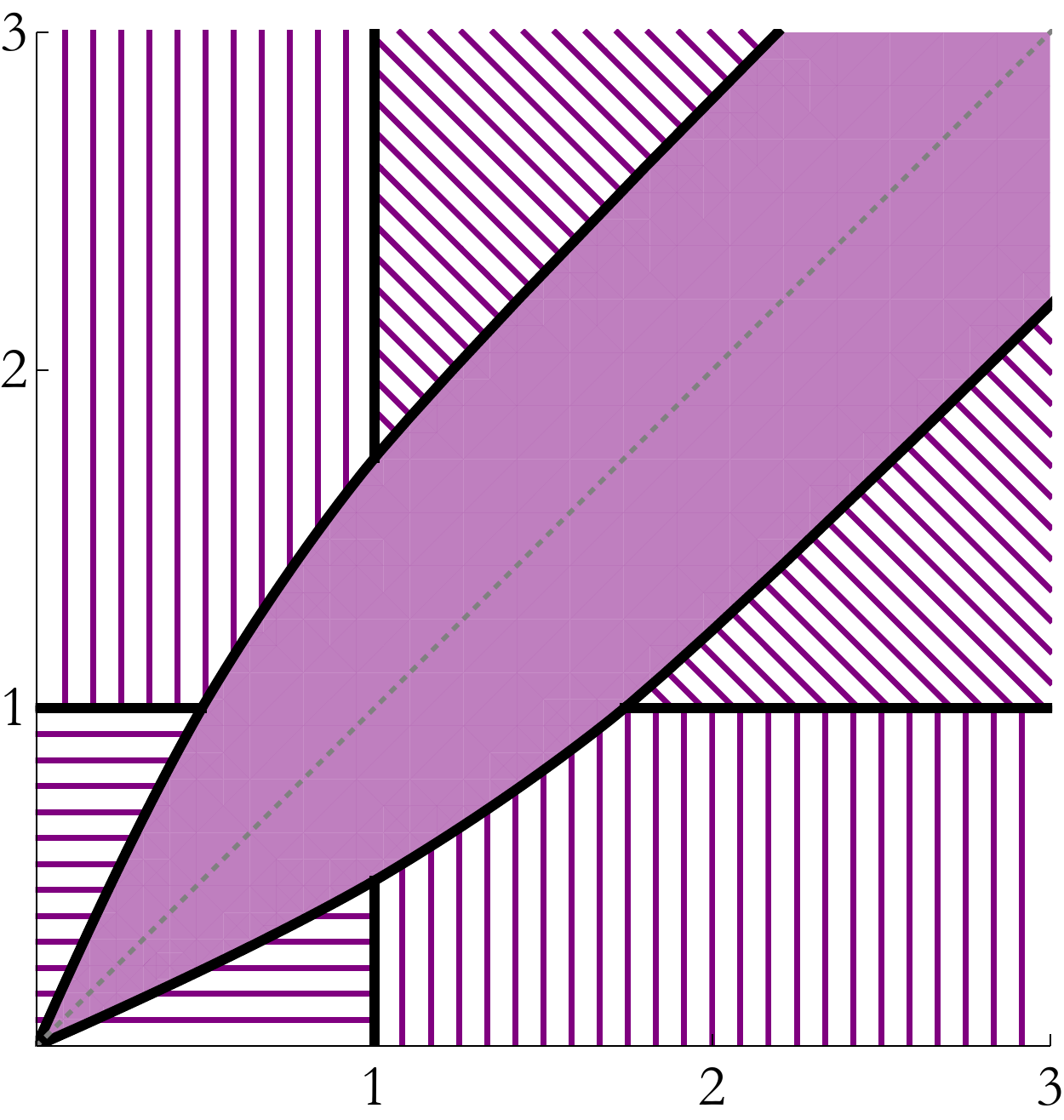} 
			\put(-185,210){$R_2/R_c$}
			\put(-30,-10){$R_1/R_c$}
			\put(-30,-35){$ $}
		\end{subfigure}
			\includegraphics[width=\textwidth]{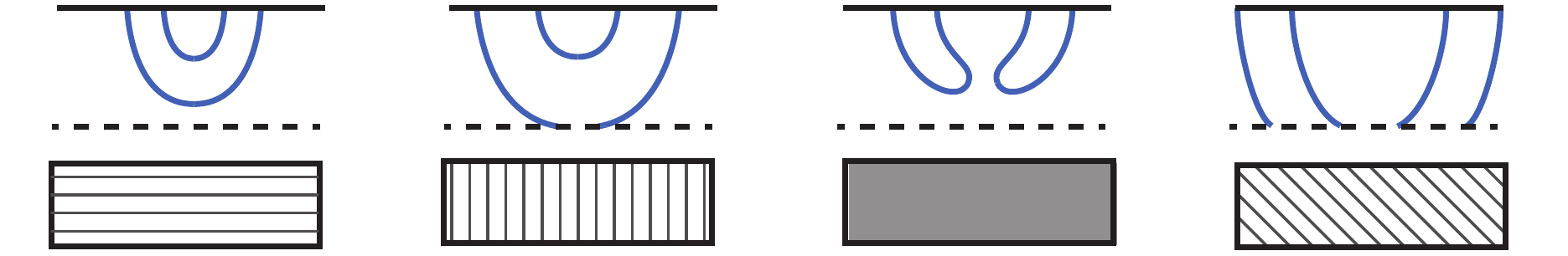} 
		\caption{\small Phase diagram stemming from the analysis of mutual information of entanglement entropy of disks in the gapped non-confining theory $\B_8^0$ (Left) and in the confining one $\Bconf$ (Right). On the axes we show the radius $(R_1,R_2)$ of the corresponding disks normalized to the value $R_c$ of the minimum radius for which the RT surface reaches the end-of-space.
		}\label{fig.mutualInfo}
	\end{center}
\end{figure}

Apart from the embeddings we just mentioned, there is another profile for the embedding that satisfies the equation of motion and is important to be taken into account. We depict this in Fig.~\ref{fig:disk_conf}(c). It naturally emerges when one imposes the boundary conditions
\begin{equation}
\label{eq:embeddding_teo_disks}
r(\rho_*) =  r_*\qquad , \qquad  \dot r(\rho_*) = 0 \ ,
\end{equation}
with $r_*\neq r_s$ and $\rho_*\neq 0$. Although the picture is self-explanatory, the way in which the equation of motion with the boundary conditions (\ref{eq:embeddding_teo_disks}) is solved is slightly involved and hence explained in detail in Appendix~\ref{ap:disk}. The relevant outcome is that this embedding is attached to two disks at the boundary, thus allowing us to compute mutual information as we did in \eqref{eq:mutual_info}. We can therefore map out the phase diagram in terms of the two radii $(R_1,R_2)$ where the mutual information (\ref{eq:mutualpreinfo}) associated with disk configurations vanishes and showing the regions where each configuration is dominant. This phase diagram is plotted in Fig.~\ref{fig.mutualInfo}.

\section{Limiting cases}\label{sec:limiting}

Having discussed the generic features and compared the confining and the non-confining theories, we now narrow down the scope and discuss specific cases close to the limiting values for the parameter $b_0$.
As described in Section \ref{sec:background}, these limiting values of $b_0$ are interesting because they lead to radically different IR dynamics. In this Section we want to study these different limits from the point of view of information probes and illustrate their power in revealing interesting physics.

\subsection{Quasi-confining regime}
\begin{figure}[t]
	\begin{center}
		\begin{subfigure}{.65\textwidth}
	\includegraphics[width=\textwidth]{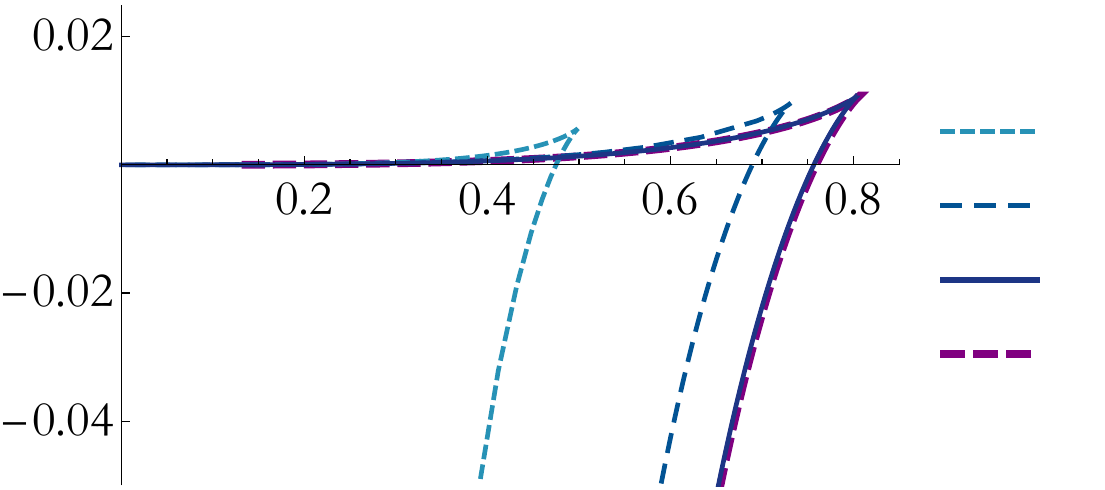} 
			\put(-245,130){$\overline S^{\text{reg}}$}
			\put(-60,55){$\overline l$}
			\put(-7,88){$b_0 = 0.6835$}
			\put(-7,69){$b_0 = 0.9201$}
			\put(-7,50){$b_0 = 0.9972$}
			\put(-7,31){$b_0 = 1 $ $(\Bconf)$}
		\end{subfigure}\vspace{7mm}
		\begin{subfigure}{0.65\textwidth}
			\includegraphics[width=\textwidth]{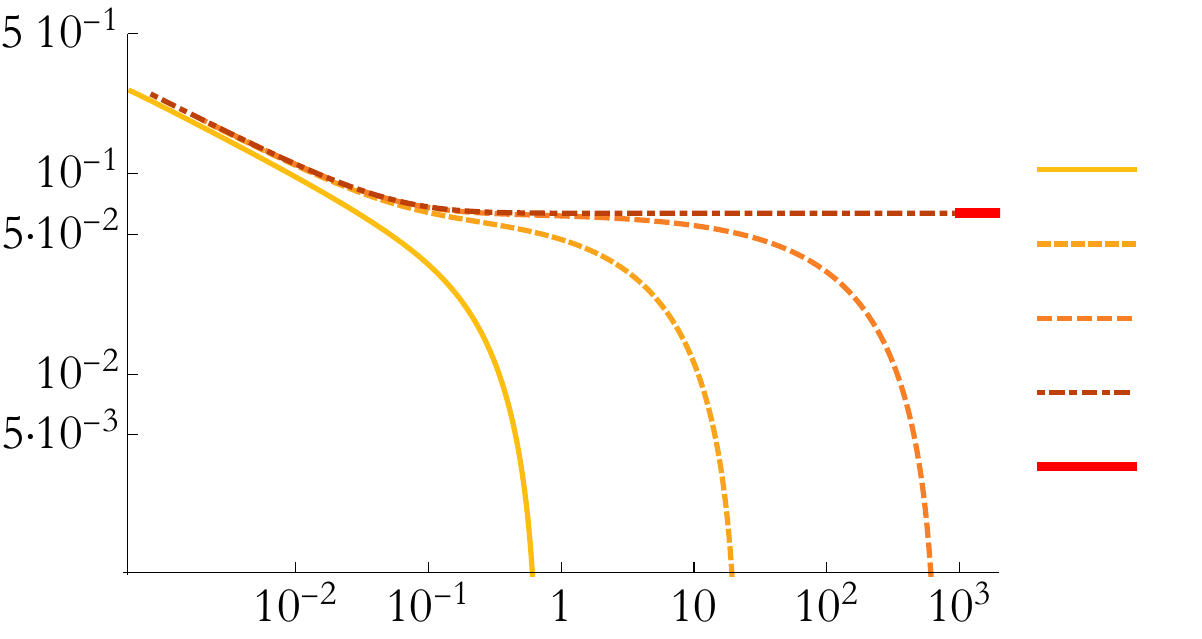} 
			\put(-240,145){$\overline S^{\text{reg}} \cdot \overline l$}
			\put(-50,-5){$\overline l$}
			\put(-7,106){$b_0 = 0.1914$}
			\put(-7,88){$b_0 = 0.0604$}
			\put(-7,70){$b_0 = 0.0191$}
			\put(-7,52.5){$b_0 = 0 $ $(\B_8^\infty)$}
			\put(-7,36){OP | CFT}
		\end{subfigure}
		\caption{\small  (Bottom) Entanglement entropy of a single strip (multiplied by width $l$) as we dial the parameter $b_0$ to zero. In all cases for non-vanishing $b_0$ we find that this quantity vanishes, contrary to the case $b_0=0$ where we find precisely the CFT result. Notice, that closer the RG flow passes the OP fixed point, the wider the entanglement plateau, before the entanglement entropy plunges.  (Top) Entanglement entropy of a strip as we increase $b_0$ and hence creeping towards the confining theory. Note that for the penultimate value for $b_0$, the curve is overlapping with that of $\Bconf$.
		}\label{fig.limiting_EEstrip}
	\end{center}
\end{figure}

\begin{figure}[t]
	\begin{center}

		\begin{subfigure}{.45\textwidth}
			\includegraphics[width=\textwidth]{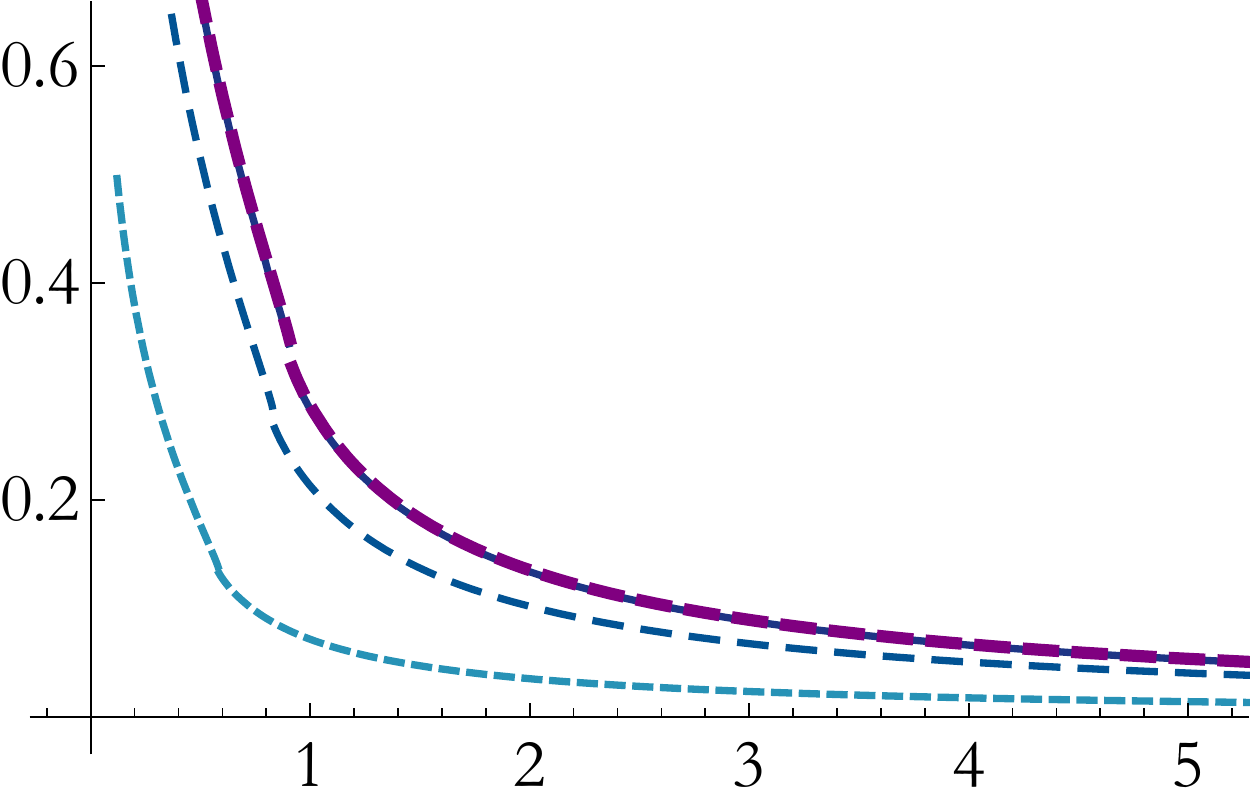} 
			\put(-185,135){$\overline  \ffunc_D(R)$}
			\put(-10,-10){$\overline R $}
		\end{subfigure}
		\hfill
		\begin{subfigure}{0.45\textwidth}
			\includegraphics[width=\textwidth]{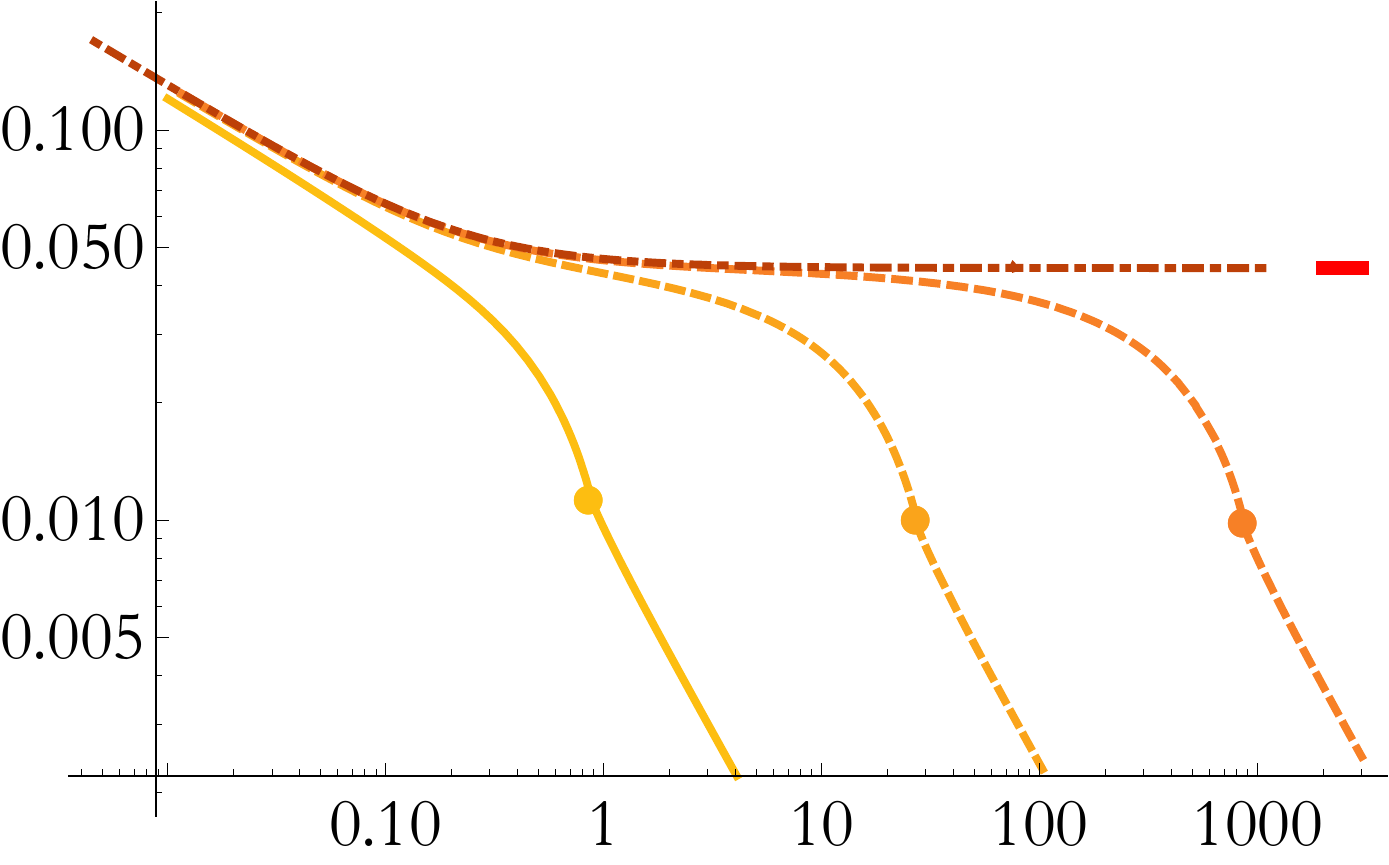} 
			\put(-185,135){$\overline  \ffunc_D(R)$}
			\put(-10,-10){$\overline  R$}
		\end{subfigure}
		\caption{\small (Right) Function $\overline \ffunc_D$ from the entanglement entropy of a disk as a function of its radius $R$. Dots stand for the transition of the embedding, namely those represented in Fig.~\ref{fig:disk_conf}(a) to \ref{fig:disk_conf}(b). (Left) Same quantities in linear scale as we approach the confining theory. In both cases, when the transition happens, there is a visible change in the behavior of the curves. This signals the proximity of the IR gapped phase. See legends in Fig.~\ref{fig.limiting_EEstrip}.
		}\label{fig.limiting_FofRdisk}
	\end{center}
\end{figure}

\begin{figure}[t]
	\begin{center}
		\begin{subfigure}{.45\textwidth}
			\includegraphics[width=\textwidth]{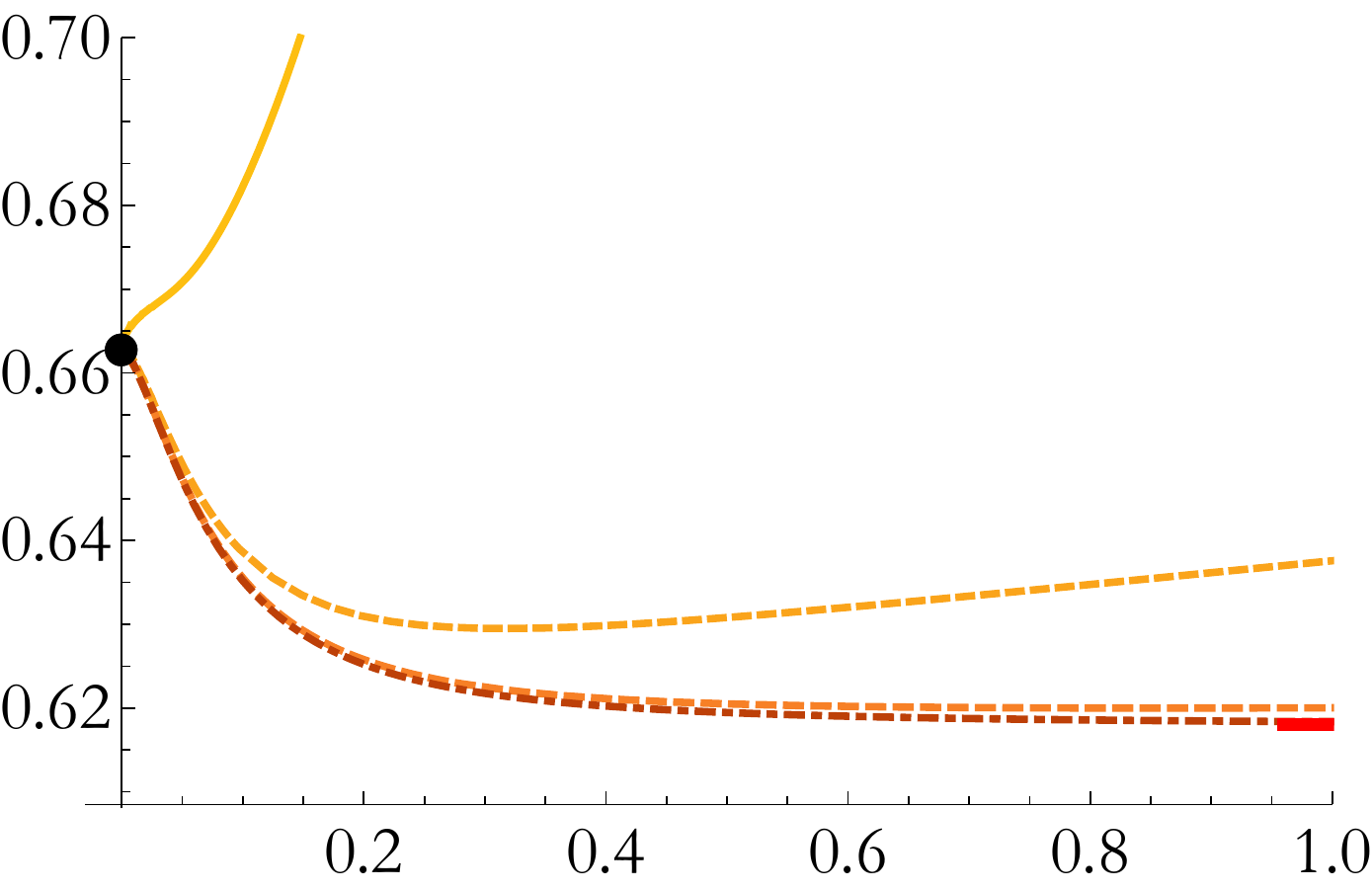} 
			\put(-185,135){$s/l$}
			\put(-30,-10){$\overline l$}
		\end{subfigure}\hfill
	\begin{subfigure}{0.45\textwidth}
			\includegraphics[width=\textwidth]{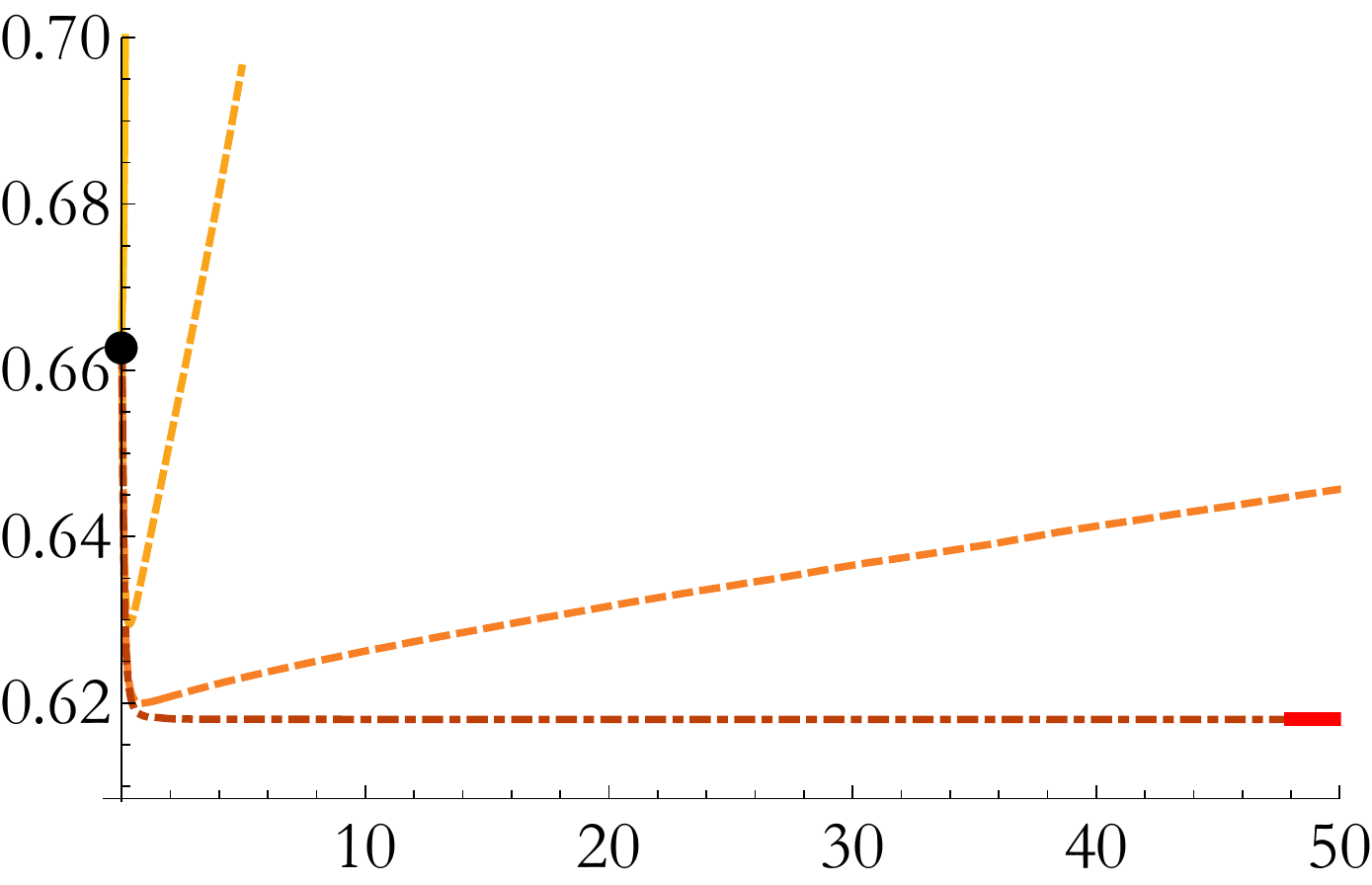} 
			\put(-185,135){$s/l$}
			\put(-30,-10){$\overline  l$}
		\end{subfigure}
		\caption{\small Ratio of the values of the width of strips $l$ which are separated by a distance $s$ at the point, where mutual information vanishes. Right plot extends the range of the left plot to larger strip widths. Notice that the numerics match analytic results both for small $l$ at the UV (\ref{eq:mutualD2}) and large $l$ at the IR (\ref{eq:mutualCFT}). See legends in Fig.~\ref{fig.limiting_EEstrip}.
		}\label{fig.golden}
	\end{center}
\end{figure}

Let us start with the case $b_0=1$, which corresponds to the confining theory $\Bconf$. If we aim to understand this solution as a limit of the $\B_8$ family of theories, the charge $Q_k$ -- related to the CS level by \eqref{eq:gauge_parameters} -- must be rescaled. This is needed in order to obtain vanishing Chern-Simons level \cite{Faedo:2017fbv,Elander:2018gte}. After performing this rescaling, in the limit $b_0\to 1$, all quantities will approach the values that can be obtained directly in the case $\Bconf$. This fact is illustrated in Fig.~\ref{fig.limiting_EEstrip} (Top), where we find that the entanglement entropy of the strip smoothly approaches the one for the confining case as $b_0$ is gradually raised towards unity.
We also show the same expected behavior for the number of degrees of freedom, {\emph{i.e.}}, for $\ffunc_D$ function in Fig.~\ref{fig.limiting_FofRdisk} (Left) for disk configurations.

\subsection{Quasi-conformal regime}

On the opposite limit, for $b_0=0$ the ground state RG flow, denoted by $\B_8^\infty$, ends at a fixed point, as indicated by the leftmost arrow in Fig.~\ref{fig:triangle}. Flows with $b_0 > 0$ approach the fixed point but never reach it. We will investigate the imprints that this passage close to the conformal fixed point leaves on information theoretic quantities. 

To start with, the OP conformal point has the standard expression for the finite part of the entanglement entropy, which is readily available by analytic methods. For strips, the full expression reads
\begin{equation}\label{eq:OPconformal}
\Delta S_{\text{OP}}(l) =  - \ \frac{9 \ \lambda\ L_y\ V_6}{2^8 \pi^4}\cdot \frac{|k|\left(\bar{M}^2+2|k|N_c\right)}{N} \ 
\times \
\frac{200\ \pi^3}{729\cdot \Gamma\left[\frac{1}{4}\right]^4}\sqrt{\frac{5}{3}} \cdot \frac{1}{l} \ .
\end{equation}
Note that $ \Delta S_{\text{OP}}(l) \cdot l $ is a constant depending on the number of degrees of freedom and we denote this as a plateau in the following.
As a consequence, flows with small $b_0$ pass close to the CFT also induce plateaux, which are wider the closer we are to the fixed point. Interestingly, it has been noted that 
the proximity to fixed points causes ``walking regime'', {\emph{i.e.}}, a regime of energy where some dimensionless quantities are approximately constant, even when such fixed points reside in the complex plane \cite{Gorbenko:2018ncu,Faedo:2019nxw}.

The entanglement entropy of a single strip times its width for different values of $b_0$ approaching zero is shown in Fig.~\ref{fig.limiting_EEstrip} (Bottom). Note again that we plot the rescaled quantities as defined in Appendix~\ref{ap:strip}. It is easy to see that the closer we are to $b_0=0$, the wider the range where $ \Delta \overline S \cdot \overline l$ traces the conformal value given by (\ref{eq:OPconformal}). As soon as the quasi-conformal regime is departed, the curve abruptly decreases towards zero.

The near-proximity of the conformal fixed point will leak to the behavior of many other quantities. Perhaps most prominently this fact is captured by the $c$-functions. Indeed, let us consider the function that measures the degrees of freedom for the disk configurations. Quasi-conformal regime is clearly visible in Fig.~\ref{fig.limiting_FofRdisk} (Right). Note, however, that $\overline  \ffunc_D$ vanishes asymptotically: the way we see this in that plot is that, once the embedding reaches the IR of the geometry, the behavior of $\overline  \ffunc_D$ rapidly renders into a straight line in the log-log plot. From its slope we can conclude that, for large values of $R$, the function $\overline \ffunc_D$ decreases  as $1/R$ for any $b_0$, which agrees with \cite{Klebanov:2012yf}.

Finally, we would like to show another quantity where quasi-conformal regime becomes manifest, arising from the computation of mutual information. Recall from above the computation of the mutual information between two parallel strips of the same width $l$ which are separated by a distance $s$ in \eqref{eq:mutual_info}. The values $(s,l)$ for which it vanishes are represented in plot Fig.~\ref{fig.mutualInfostrips}. Interestingly, this critical ratio in a CFT is universal and given by the golden ratio \cite{Ben-Ami:2014gsa,Balasubramanian:2018qqx}
\begin{equation} \label{eq:mutualCFT}
\frac{s}{l}\Big |_{\text{CFT}}= \varphi^{-1} = \frac{-1+\sqrt{5}}{2} \approx 0.618 \ .
\end{equation} 
Moreover, in the UV this ratio is also fixed in all cases by D2-brane asymptotics \cite{vanNiekerk:2011yi}, leading to 
\begin{equation}\label{eq:mutualD2}
\frac{s}{l}\Big|_{\text{D2}} = -1+\sqrt{1+\beta}\approx 0.663 \ ,
\end{equation} 
where $\beta$ is the (unique) real root of the polynomial 
\begin{equation}
 b(x)=64 x^{11}-64 x^{10}+16 x^9-400 x^8+x^7+191 x^6+768 x^5+744 x^4-192 x^3-704 x^2-1024 x-512 \ .
\end{equation}
As we analyze the mutual information $I(s,l)$, and in particular the boundary where it vanishes and defines the critical ratio $s/l$ for different theories, we find rich behavior revealed in Fig.~\ref{fig.golden} that can be interpreted as follows. Let us consider small $b_0$ values, so that the flows approach the OP fixed point. At the UV, {\emph{i.e.}} for small widths $l$, the critical ratio $s/l$ starts from the UV value (\ref{eq:mutualD2}), decreasing and developing a global minimum before increasing again towards IR (large $l$). For even smaller values of $b_0$ the critical $s/l$ curve can get arbitrarily close to the CFT value (\ref{eq:mutualCFT}), eventually diverging from it. However, only in the strict $b_0=0$ case, corresponding to $\B_8^\infty$ will we reach (\ref{eq:mutualCFT}) in the asymptotic IR regime. Our results are suggestive that the CFT value acts as a lower bound on critical $s/l$ and it would be interesting to understand the reason behind this.

\section{Conclusions and Discussion}\label{sec:conclusions}

The main message of this paper is that holographic entanglement entropy and Wilson loops cannot be considered interchangeably as mediators of the fact whether the theory is confining. Whereas the quark-antiquark potential is sensitive to the fact that the flux tube between two infinitely massive quarks cannot break apart, entanglement entropy seems to be signaling the presence of a mass gap, capping off the flow of information. An observation worth highlighting, in the current context, is that the entanglement measures are insensitive to the Chern-Simons interactions, which may be of relevance. This aspect deserves to be more properly understood. To this end, we hope to make closer contact with recent important studies of entanglement entropies in $(2+1)$-dimensional Chern-Simons field theories \cite{Agarwal:2016cir}. 

\begin{figure}[t]
	\begin{center}
		\includegraphics[width=.7\textwidth]{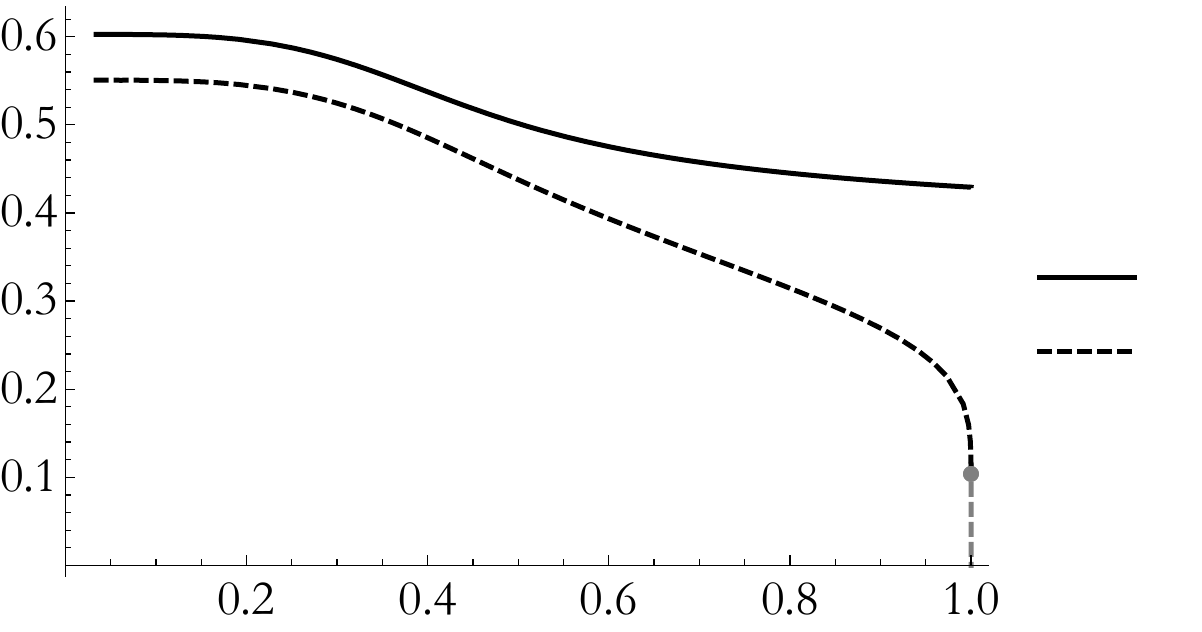} 
		\put(-7,84){$l_c/l_{\text{qq},c}$}
		\put(-7,65){$l_t/l_{\text{qq},t}$}
		\put(-45,15){$b_0$}
		\caption{\small  Ratios between scales stemming from quark-antiquark potentials and entanglement entropy of strips as a function of $b_0$. The solid curve is the ratio between the critical lengths where the disconnected strings and $\sqcup$-RT-surface become dominant for the Wilson loop and for entanglement entropy of a strip, respectively. The dashed curve stands for the ratio between the turning points of hanging string configurations and the RT embeddings. The last part of this curve, drawn fainter and below the gray dot, stands for extrapolation.}
		\label{fig:compare_scales}
	\end{center}
\end{figure}

A natural question to address is how the scales that we identify in entanglement entropy measures relate to the physical scales given by the mass gap.\footnote{We thank Anton Faedo for bringing up this issue.} For that, one possibility would be to compare entanglement entropy scales to the masses of the lightest states in the spectrum of spin-2 and spin-0 particles obtained in \cite{Elander:2018gte}. It is not completely clear, however, how to make this comparison and it is furthermore not evident if the states found in \cite{Elander:2018gte} actually are the lightest ones. In the paper at hand, we are content with comparing to the scales that arise from quark-antiquark potentials. Taking Figs.~\ref{fig.WilsonLoop} and \ref{fig.EEplotsStrip} into consideration, there are several scales that could be considered. The first one is the one we denoted by $l_c$, the width of the strip at which the $\sqcup$ configuration becomes preferred. This scale roughly indicates above which width there is some region (in the middle) of the strip which is not entangled with the complement. It is natural to compare this scale to the length at which the fluxtube between quarks breaks apart, which sets the length above which the quarks become screened by CS interactions. It is therefore expected to be related to some correlation length of the ambient field theory. 
Taking into account that the quark-antiquark potential in \cite{Faedo:2017fbv} (see also Fig.~\ref{fig.WilsonLoop}~(Left)) is given by the difference between the energy of the meson and that of a disconnected pair of quarks, the critical length is one where the quark-antiquark potential vanishes.

In Fig.~\ref{fig:compare_scales} we depict the ratio set by the entanglement measure to the one resulting from quark potential. We find that this ratio is around $l_c/l_{\text{qq},c}\approx 0.5$ for all theories. The fact that both scales are of the same order is a manifestation of the fact that the correlation length induced by the screening of the quark is linked with the separation between entangled states. An interesting question is in how general this statement holds. One can address this in other theories possesing phase transitions for entanglement measures by extracting the corresponding correlation lengths, and comparing them to the critical length resulting from string breaking.

There is another useful comparison to be discussed. The turning points of the RT surfaces and the string embeddings will give rise two additional numbers. We refer to them as $l_t$ and $l_{\text{qq},t}$, respectively. These are defined as the locii of the maxima as depicted in Fig.~\ref{fig.WilsonLoop}~(Left) and Fig.~\ref{fig.EEplotsStrip}, respectively. As one dials $b_0\to 1$, {\emph{i.e.}}, we approach the confining $\Bconf$, the value of $l_{\text{qq},t}$ diverges; see Fig.~8~(Right) in \cite{Faedo:2017fbv}. As shown in Fig.~\ref{fig.limiting_EEstrip}~(Top), $l_{t}$ however saturates to the value given by the turning point of the confining theory. This causes the ratio $l_{t}/l_{\text{qq},t}$ to vanish in the limit $b_0\to 1$. This yields another  manifestation that entanglement entropy measures are not sensitive to confining dynamics.

A motivation for the claim that entanglement entropy probes confinement in \cite{Klebanov:2007ws} was the similarity between the phase transition from connected to disconnected configuration of strip entanglement entropy and the first-order finite temperature deconfinement transitions which is typically found in gravitational duals of confining gauge theories. We feel that this proposal should be viewed with caution. To be concrete, although all the $\B_8$ theories studied in this paper presented the same qualitative behavior as far as entanglement entropy is considered (except for $\B_8^\infty$), their finite temperature physics appears richer. In \cite{Elander:2020rgv} the thermodynamic phase diagram was mapped out with lavish transitions of first and second order, including a triple point and a critical point. It would be interesting to extend our analysis to finite temperature and similarly map out the phase diagram following from entanglement thermodynamics. It would be particularly compelling to make a thorough study of the entanglement phase diagram around the thermodynamic critical point of these theories. Such a top-down computation could further be compared to the analogous computation carried out in a holographic model of QCD \cite{Knaute:2017lll}.

Since the entanglement entropy does not seem to single out confining geometries from those that are not, one could raise to concern if some other set of boundary data would be better suited to bulk reconstruction. A particularly compelling option would be to lean on Wilson loops \cite{Jokela:2020auu,Hashimoto:2020mrx}, for example. In fact, in \cite{Hashimoto:2020mrx} (see also \cite{Sonnenschein:1999if}) it was argued that the linear quark-antiquark potential necessary leads to an ``IR bottom'' in the in deep interior of bulk spacetime. This, however, is not yet completely satisfactory due to a technical reason: having an regular ``IR bottom'' (typically realized by cigar-like geometries) is not in one-to-one correspondence with linear potential. This statement follows from the models we discussed. Then, the obstacle present at finite temperature is faced here as well, whence the hanging string breaks due to pair creation and therefore reconstruction of the IR part of the geometry is not possible. We feel that these issues can be resolved upon closer inspection and hope to address them in future works. 

A natural extension of our work would be to consider multiparty entanglement, between strips \cite{Ben-Ami:2014gsa}, or disks, for example, and analyze more refined probes of information flows. It would be particularly interesting to analyze how the extensivity of the mutual information \cite{Balasubramanian:2018qqx} behaves under the RG flow in our family of geometries and whether some intermediate length scales are more super-extensive than the CFT at the OP fixed point. A further investigation of entanglement measures for mixed states \cite{Takayanagi:2017knl,Kudler-Flam:2018qjo,Tamaoka:2018ned,Jokela:2019ebz,Lala:2020lcp,1822901}, contrary to pure states as studied in this paper, might shed more light on why entanglement wedge cross sections behave non-monotonically under RG flows \cite{Jokela:2019ebz}, specifically in confining geometries. Moreover, a particularly compelling scenario would be to add a magnetic (Kondo) impurity \cite{Benincasa:2012wu,Erdmenger:2013dpa,Erdmenger:2015spo} and study how the entanglement entropy behaves in our family of solutions, and in particular if some form of g-theorem holds. 
Finally, it would be interesting to make some contact with a phenomenon called partial deconfinement \cite{Watanabe:2020ufk,Hanada:2018zxn,Hanada:2019czd} and investigate if entanglement measures can lead to some new interesting insights.

\vspace{1.0cm}
\begin{acknowledgments}
We thank Joydeep Chakravarty, Anton Faedo, Carlos Hoyos, David Mateos, and Alan Rios for useful discussions. N.~J. has been supported in part by the Academy of Finland grant no. 1322307. J.~G.~S. acknowledges support from the FPU program, Fellowships FPU15/02551 and EST18/00331, and has been partially supported by grants no. FPA2016-76005-C2-1-P, FPA2016-76005-C2-2-P, SGR-2017-754, and by the State Agency for Research of the Spanish Ministry of Science and Innovation through CEX2019-000918-M.
\end{acknowledgments}

\appendix

\section{Details of the background solutions}\label{ap:solution}

In this appendix we review the main features of Type IIA supergravity solutions dual to the family of gauge theories we have been considering in this paper. As mentioned in Sec.~\ref{sec:background}, these solutions are regular only in eleven dimensions, we review the ten-dimensional setup instead since the geometric picture is cleaner. For the detailed discussion on eleven dimensions, see \cite{Faedo:2017fbv}.

The internal compact space is squashed $\CP^3$ and the UV geometry coincides with that induced by a stack of D2-branes. We therefore describe the geometry given the Ansatz
\begin{eqnarray}
\label{10DansatzAP}
\dd s_{\rm st}^2 &=&h^{-\frac12}\left(-\ \dd t^2 + \dd x_1^2 + \dd x_ 2^2\right)+h^{\frac12} \left({\dd r^2}+e^{2f}\dd\Omega_4^2+e^{2g}\left[\left(E^1\right)^2+\left(E^2\right)^2\right] \right)\nonumber \\
e^\Phi&=&h^{\frac14} \, e^\Lambda \,,
\end{eqnarray}
for the metric and the dilaton field. The dilaton $\Phi$ and the functions $f$, $g$, $h$, and $\Lambda$ are functions of the radial coordinate $r$. Also, the complex projective plane is seen as the coset $\rm{Sp}(2)/\rm{U}(2)$, consisting of a $\rm{S}^2$ (described by the vielbeins $E^1$ and $E^2$) fibered over $\rm{S}^4$, with metric $\dd\Omega^2_4$. A convenient choice of coordinates is the following \cite{Conde,Jokela:2012dw}. Let $\omega^i$ be a set of left-invariant one-forms on the three-sphere. The metric of the four-sphere with unit radius can be written as 
\begin{equation}
\dd\Omega_4^2\,=\,\frac{4}{\left(1+\xi^2\right)^2}\left[\dd \xi^2+\frac{\xi^2}{4}\omega^i\omega^i\right]\ ,
\end{equation}
where $\xi$ is a non-compact coordinate. Choosing $\theta$ and $\varphi$ as angles to parametrize the two-sphere, the non-trivial fibration is described by the vielbeins
\begin{eqnarray}
\label{introduced}
E^1&=&\dd \theta+\frac{\xi^2}{1+\xi^2}\left(\sin\varphi\,\omega^1-\cos\varphi\,\omega^2\right) \nonumber \\
E^2&=&\sin\theta\left(\dd\varphi-\frac{\xi^2}{1+\xi^2}\omega^3\right)+\frac{\xi^2}{1+\xi^2}\cos\theta\left(\cos\varphi\,\omega^1+\sin\varphi\,\omega^2\right) \ .
\end{eqnarray}
For our purposes, it is better to consider a rotated version of \eqref{introduced} in the four-sphere, namely
\begin{eqnarray}
\mathcal{S}^1&=&\frac{\xi}{1+\xi^2}\left[\sin\varphi\,\omega^1-\cos\varphi\,\omega^2\right] \nonumber\\
\mathcal{S}^2&=&\frac{\xi}{1+\xi^2}\left[\sin\theta\,\omega^3-\cos\theta\left(\cos\varphi\,\omega^1+\sin\varphi\,\omega^2\right)\right] \nonumber\\
\mathcal{S}^3&=&\frac{\xi}{1+\xi^2}\left[\cos\theta\,\omega^3+\sin\theta\left(\cos\varphi\,\omega^1+\sin\varphi\,\omega^2\right)\right] \nonumber\\
\mathcal{S}^4&=&\frac{2}{1+\xi^2}\,\dd\xi \ .
\end{eqnarray}
Even though ${\mathcal{S}}^i$ depend on the angles of the two-sphere, it holds that $\mathcal{S}^n\mathcal{S}^n=\dd \Omega_4^2$.
Using these vielbeins we can construct a set of left-invariant forms on the coset. This set contains the two-forms
\begin{equation}
 X_2\,=\,E^1\wedge E^2  \qquad , \qquad  J_2\,=\,\mathcal{S}^1\wedge\mathcal{S}^2+\mathcal{S}^3\wedge\mathcal{S}^4\ ,
\end{equation}
as well as the three-forms 
\begin{eqnarray}
X_3&=&E^1\wedge\left(\mathcal{S}^1\wedge\mathcal{S}^3-\mathcal{S}^2\wedge\mathcal{S}^4\right)-E^2\wedge\left(\mathcal{S}^1\wedge\mathcal{S}^4+\mathcal{S}^2\wedge\mathcal{S}^3\right)
\nonumber\\
J_3&=&-E^1\wedge\left(\mathcal{S}^1\wedge\mathcal{S}^4+\mathcal{S}^2\wedge\mathcal{S}^3\right)-E^2\wedge\left(\mathcal{S}^1\wedge\mathcal{S}^3-\mathcal{S}^2\wedge\mathcal{S}^4\right)\ .
\end{eqnarray}
These are related by exterior differentiation as  
\begin{equation}
\dd X_2\,=\,\dd J_2\,=\,X_3 \qquad , \qquad  \dd J_3\,=\,2\left(X_2\wedge J_2+J_2\wedge J_2\right)\ .
\end{equation}

Additionally, we can construct higher forms by wedging these which will also be left-invariant. Then we have the two four-forms $X_2\wedge J_2$ and $J_2\wedge J_2$, appearing in the equation above, together with the volume form $\Omega_6 = - (E_1 \wedge E_2)\wedge (\mathcal{S}^1\wedge\mathcal{S}^2\wedge\mathcal{S}^3\wedge\mathcal{S}^4)$. There are no left-invariant one- or five-forms. The fluxes can be written in terms of these left-invariant forms, since this symmetry ensures the consistency of the Ansatz. By this we mean that all the internal angles will drop from the resulting equations, leading to dependence just on the radial coordinate. Therefore, we take the following forms 
\begin{equation}\label{fluxesansatz}
\begin{array}{rclcrcl}
F_4 &=&  \dd^3x\wedge \dd(h^{-1}e^{-\Lambda})+ G_4 + B_2\wedge F_2 \qquad\qquad  & , \qquad\qquad & F_2& =& Q_k (X_2 - J_2) \\
G_4 &=& \dd(a_J J_3) + q_c\left(J_2\wedge J_2 - X_2\wedge J_2\right)\qquad\qquad & , \qquad\qquad & B_2& =& b_X X_2 + b_J J_2 \ ,
\end{array}
\end{equation}
where 
\begin{equation}\label{trunc}
a_J\,=\,\frac{e^{2g}\big( Q_kb_J-q_c\big)-2e^{2f+g-\Lambda}\big(b_J+b_X\big)+e^{2f}\Big[q_c+Q_k\big(b_X-b_J\big)\Big]}
{2\big(e^{2f}+e^{2g}\big)}\ .
\end{equation}
The parameters $Q_k$ and $q_c$ are constants which will be related to gauge theory parameters as we shall see later. Moreover, $b_J$, $b_X$, and $a_J$ are functions that depend on the radial coordinate which will satisfy first order equations as we argue next.

The system is $\NN=1$ supersymmetric and the resulting BPS equations from the supergravity action can be solved in three consecutive steps. First, we can solve the equations for the metric functions, which follow directly from \cite{Conde,Bea:2013jxa} and read
\begin{eqnarray}
\label{BPSsystem}
\Lambda'&=&2Q_k\,e^{\Lambda-2f}-Q_k\,e^{\Lambda-2g} \nonumber\\
f'&=&\frac{Q_k}{2}\,e^{\Lambda-2f}-\frac{Q_k}{2}\,e^{\Lambda-2g}+e^{-2f+g} \nonumber\\
g'&=&Q_k\,e^{\Lambda-2f}+e^{-g}-e^{-2f+g} \ .
\end{eqnarray}
Once the functions of the metric are known, one turns to solve the equations for the fluxes \cite{Faedo:2017fbv}
\begin{eqnarray}
b_X' & = & 2\,e^{-4f+2g+\Lambda} \big( q_c+2\,a_J-Q_k\,b_J\big) \nonumber \\
b_J'&=&e^{-2g+\Lambda}\Big[ Q_k\left(b_J-b_X\right) +2\,a_J-q_c\Big]\ , \\
\end{eqnarray}
where $a_J$ is given by \eqref{trunc}. The warp factor $h$ follows by direct integration of
\begin{eqnarray}
\left(e^\Lambda\,h\right)' &=& -e^{2\Lambda-4f-2g}\Big[Q_c+Q_k\,b_J \left(b_J- 2 b_X\right)+ 2 q_c \left(b_X- b_J\right) + 4 a_J\left(b_X+ b_J\right)\Big] \ .\qquad 
\end{eqnarray}
Here, $Q_c$ appears as an integration constant. Together with $Q_k$ and $q_c$ it relates to gauge theory parameters via the quantization of the Page charge \cite{Marolf:2000cb}, which leads to \cite{Hashimoto:2010bq,Faedo:2017fbv}:
\begin{equation}\label{eq:gauge_parameters}
Q_c = 3\pi^2 \ls^5 g_s N,\qquad Q_k = \frac{\ls g_s}{2} k, \qquad q_c =\frac{3\pi \ls^3g_s}{4} \bar M =  \frac{3\pi \ls^3g_s}{4} \left(M-\frac{k}{2}\right)\ .
\end{equation}
In the previous expression, $N$ is the number of D2-branes and matches the rank of the gauge group on the field theory. The parameter $Q_k$ gives rise to D4-brane charge and $k$ is consequently expected to be the CS level of the dual gauge theory. Finally, $q_c$ is related to the number of fractional D2-branes that is needed to obtain a regular geometry \cite{Cvetic:2001ma}, where $M$ represents the shift in the gauge group due to fractional branes \cite{Aharony:2008gk}. In addition, notice that the string length $\ls$ and the string coupling constant $g_s$ are related to the gauge coupling constant via $\lambda =\ls^{-1} g_s N$, which is dimensionful $\sim$(length)$^{-1}$ in three dimensions.

For non-vanishing CS level $|Q_k|\neq0$ the dependence on the different charges can be factored out of the equations by writing them in terms of dimensionless functions $\FF$, $\GG$, $\BB_X$, $\BB_J$, and $\mathbf{h}$ defined through\footnote{A similar factorization of charges with an appropriate radial coordinate can be performed when $|Q_k|=0$, see \cite{Faedo:2017fbv}.}
\begin{equation}\label{eq:dimlessfunctions}
\begin{aligned}
e^f &=  |Q_k|\ e^\FF\, , \qquad \qquad  e^g = |Q_k|\ e^\GG\,,  \qquad\qquad h= \frac{4q_c^2 +3 |Q_k| Q_c}{|Q_k|^6} \ \mathbf{h} \\
b_J &= - \frac{2q_c}{3 |Q_k|} - \frac{\sqrt{4 q_c^2 + 3 |Q_k| Q_c}}{3 |Q_k|}\ \BB_J\,, \qquad\quad b_X =  \frac{2q_c}{3 |Q_k|} + \frac{\sqrt{4 q_c^2 + 3 |Q_k| Q_c}}{3 |Q_k|}\ \BB_X \ ,
\end{aligned}
\end{equation}    
together with the dimensionless radial coordinate
\begin{equation}\label{ucoord}
u\,=\,\frac{|Q_k|}{r} \ .
\end{equation}
In \cite{Faedo:2017fbv} a different radial coordinate $y$ was used. It is defined via 
\begin{equation}\label{eq:coordy}
\dd r = -\frac{P}{Q_k\ v\ (1-y^2)} \dd y
\end{equation}
where $v$ and $P$ are solutions of the following differential equations:
\begin{equation}
2\ (1-y^2)\ \frac{\dd v}{\dd y} = y\ v +2 \qquad, \qquad\frac{1}{P}\ \frac{\dd P}{\dd y} = \frac{v+1}{v\ (1-y^2)}\ .
\end{equation}
In terms of $y$, all the functions of the metric except $h$ are expressed analytically.
Nevertheless, we preferred to use \eqref{ucoord} because it simplifies the UV expansion, which will be crucial to renormalize the entanglement entropy of disks. In fact, the UV expansion of our system can be directly read off from \cite{Elander:2020rgv}, where finite temperature states of these theories where constructed. Setting the parameter that renders the temperature $b_5$ there to zero, one arrives at
\begin{equation} 
\begin{aligned}
e^{2\FF} &= \frac{1}{2 u^2} \left[ 1-2u-4u^2-6u^3+\parent{\frac{77}{4}+2f_4}u^4+\parent{\frac{65}{2}-2f_4+2f_5}u^5+\OO(u^{6}) \right] \\[2mm]
e^{2\GG} &= \frac{1}{4u^2} \left[1-4u-4u^2-\parent{-\frac{109}{2}-4f_4}u^4 + \parent{\frac{821}{2}+14f_4+2f_5}u^5+\OO(u^{6})\right] \\[2mm]
e^{\Lambda} &= 1 - 4u^2 -16 u^3 -48u^4+\parent{-\frac{71}{10}+\frac{6}{5}f_4+2f_5}u^5+\OO(u^{6}) \\[2mm] \nonumber
\end{aligned}
\end{equation}
\begin{equation}\label{eq:solutionUV}
\begin{aligned}
\BB_J &= b_0\left[ 1+4u+8u^2-16u^3+\frac{b_4}{b_0} u^4 + \OO(u^5) \right] \\[2mm]
\BB_X&= b_0\left[1+4 u+12 u^2+32 u^3- \left(64 + \frac{b_4}{2 b_0} \right) u^4+O\left(u^5\right)\right] \\[2mm]
\textbf{h}&= \frac{16}{15}  (1-b_0^2)\ u^5 \left[1+\frac{20(1-2b_0^2)}{3(1-b_0^2)}u+\frac{4(581b_0^2-201)}{21(-1+b_0^2)}u^2+\OO(u^3)\right] \ .
\end{aligned}
\end{equation}
Note that there are several undetermined parameters in this expansion, some of which are going to be fixed now by supersymmetry. For example, $f_5$ is fixed in terms of $f_4$,
\begin{equation}
f_5 = -3f_4 - \frac{411}{4} .
\end{equation}
Similarly, there are additional higher order undetermined terms present in \cite{Elander:2020rgv} which we can fix here in terms of $f_4$, $b_0$, and $b_4$ by supersymmetry:
\begin{eqnarray}
b_6 &=& \frac{2}{5} \, \left(\, b_0\,  \left(\, 200 \, f_4\, +\, 709\, \right)\, +\, 98 b_4\, \right) \nonumber\\
b_9 &=& \frac{1}{140} \left(\, 28 b_4 \, \left(\, 104 f_4\, +\, 26441 \, \right)\, +\, b_0\,  \left(\, 3008 f_4^2\, +\, 1202736 f_4\, -\, 1641593\, \right)\right) \nonumber \\
f_{10} &=& - \frac{1}{1120}\, \left( \, 6472 f_4^2\, +\, 1028728 f_4\, +\, 33459213\, \right) \ .
\end{eqnarray}
So in the end there are three UV parameters, namely $b_0$, $b_4$, and $f_4$. 
On one hand, $b_0$ is the parameter distinguishing the different solutions appearing in Fig.~\ref{fig:triangle}. On the other hand, once $b_0$ is fixed, $b_4$ and $f_4$ are fixed by requirement of regularity at the IR. 

Let us mention that when the coordinate $y$ given by \eqref{eq:coordy} is considered, a parameter $y_0$ arises, which is its value at the end of space. The parameter $y_0$ is in a one-to-one (numerical) correspondence with $b_0$ and was the one used in \cite{Faedo:2017fbv} for labeling the $\B_8$ theories. Even though $y_0$ does not appear in our analysis, let us relate it with the parameters we have just found for the sake of completeness. For that we give the expressions for $f_4$ and $b_4$ in terms of $y_0$, so that the interested reader has the precise link to \cite{Faedo:2017fbv}. Concretely, \footnote{The parameter $\beta_4$ here was called $b_4$ in \cite{Faedo:2017fbv}.}
\begin{equation}
f_4 = \frac{1}{24}(- 423 -w_\pm^4(y_0))\,,\qquad b_4 = \frac{ \ b_0 \ \beta_4 }{2} \ ,
\end{equation}
where the subscript in $w_\pm(y_0)$ consistently refers to either $\B_8^+$ (when $b_0\in(0,2/5)$) or $\B_8^-$ (when $b_0\in(2/5,1)$) and their expressions in each case are given by \footnote{Note that $w_\pm(y_0)$ are related to $P_0^{\pm}$ from \cite{Faedo:2017fbv} via $4P_0^{\pm} = |Q_k|^2 w_\pm (y_0)$.}
\begin{eqnarray}
 w_+(y_0) &=& \frac{\Gamma\left[\frac{1}{4}\right]^2}{\sqrt{8\pi}} + 2 \left({1-y_0^2}\right)^{\frac{1}{4}}-y_0 \, _2F_1\left(\frac{1}{2},\frac{3}{4};\frac{3}{2}; y_0^2\right) \\
 w_-(y_0) &=& \sqrt{8 \pi }\cdot \frac{\Gamma
	\left(\frac{5}{4}\right)}{\Gamma \left(\frac{3}{4}\right)}+2\left(y_0^2-1\right)^{\frac{1}{4}} -\frac{2}{\sqrt{y_0}} \,_2F_1\left(\frac{1}{4},\frac{3}{4};\frac{5}{4};\frac{1}{y_0^2}\right)\,.
\end{eqnarray}

Apart from the UV expansion, we also need the IR expansion of the metric when setting the boundary conditions for the RT embeddings reaching the bottom of the geometry. After a straightforward calculation, the first few orders of the IR expansions for $b_0\in (0,1)$ read:
\begin{eqnarray}\label{eq:IRexpansionmetric}
e^{2\FF} &=& \frac{f_s^2}{u_s^2} \, (u_s-u)\  +\  \left(\frac{3}{u_s^4}+f_s^2 \lambda _s\right)\,  (u_s-u)^2  \nonumber\\[0mm]
&& +\  \frac{
	3 f_s^4 \lambda _s^2 u_s^8 \ +\  10 f_s^2 \lambda _s u_s^4\ + \ 8 f_s^2  u_s \ +6}{3f_s^2u_s^6}\, (u_s-u)^3\ 
+\ \OO(u_s-u)^4 \nonumber\\[3mm]
e^{2\GG} &=&\frac{1}{u_s^4} \,  (u_s-u)^2 \ +\ \frac{2}{u_s^6} \left(u_s-\frac{1}{f_s^2}\right) \, (u_s-u)^{3} \ +  \  \OO(u-u_s)^4\ \nonumber\\[3mm]
e^{\Lambda} &=& \frac{1}{u_s^2}\, (u_s-u) \ +\  \lambda_s\,  (u_s-u)^2  \nonumber\\[0mm] 
&&  +\ \frac{\ 3 f_s^4 \lambda _s^2 u_s^8 \ + \ 4 f_s^2 u_s \left(\lambda _s u_s^3 \ - 1\right) \ + \ 9}{3 f_s^4 u_s^6} \, (u-u_s)^3\ +\ \OO(u_s-u)^4 \nonumber\\[3mm]
\BB_J &=& 1\ -\ \frac{u_s-u}{f_s^2u_s^2} \ +\  \frac{2-f_s^2u_s}{f_s^4u_s^4}\, (u_s-u)^2 \nonumber\\[0mm]
&& -\ \frac{2 f_s^2 u_s \left(\lambda _s u_s^3-31\right)+15 f_s^4 u_s^2+72}{15 f_s^6 u_s^6}\ (u_s-u)^3\ +\ \OO(u_s-u)^4 \nonumber\\[3mm]
\BB_X &=& 1-\frac{3}{f_s^4u_s^4}\, (u_s-u)^2 \ +\ \frac{2 \left(f_s^2 u_s \left(\lambda _s u_s^3-4\right)+9\right)}{f_s^6 u_s^6}\,  (u_s-u)^3\  +\ \OO(u_s-u)^4 \nonumber\\[3mm]
\mathbf{h} &=& h_{\text{IR}}\cdot\frac{1}{u_s-u} - \left(h_{\text{IR}} \lambda _s u_s^2+\frac{7}{3 f_s^8} \right)\nonumber \\[0mm]
&& + \ \frac{28 u_s^2 \left(f_s^2 u_s \left(\lambda _s u_s^3-1\right)+4\right)-3 h_{\text{IR}} f_s^6 \left(4 f_s^2 u_s
	\left(\lambda _s u_s^3-1\right)+9\right)}{9 f_s^{10} u_s^4} \, (u_s-u)  \nonumber\\[0mm]
&&  + \  \OO(u_s-u)^2
\ .
\end{eqnarray}
For $b_0=0$ the corresponding expansion is not needed in our paper, since the RT surface will not reach the bottom of the geometry. For the confining $b_0=1$ theory, for which the coordinate $u$ is not valid, an alike IR expansion can be found. It follows directly from the analytic expression of the solution \cite{Faedo:2017fbv,Cvetic:2001ma}, and we omit it here for the sake of brevity. Moreover, note that in \eqref{eq:IRexpansionmetric} we found four IR parameters which are not fixed by regularity in the IR expansion. For $u_s$, $f_s$ and $\lambda_s$ it is easy to find expressions as a function of the parameter $y_0$ from \cite{Faedo:2017fbv}:
\begin{eqnarray}
f_s^2 &=& \frac{ (1+y_0)^{\frac{3}{4}}}{\left(\pm (1-y_0)\right)^{\frac{1}{4}}}\cdot \frac{w_\pm(y_0)}{2}\,, \qquad
\lambda_s = \frac{1}{u_s^3} \ \pm \ \frac{2y_0^2+2y_0-4}{w_\pm(y_0) u_s^4\left(\pm(1-y_0^2)\right)^{\frac{3}{4}}}\,, \\ [2mm]
u_s &=& \frac{6}{w_\pm(y_0)}\left[-2^{3/4} \left(y_0\pm 1\right)^{\frac{3}{4}} \, _2F_1\left(\frac{1}{4},\frac{3}{4};\frac{7}{4};\frac{1\pm y_0}{2} \right)+3\frac{
	\left(1+y_0\right){\frac{3}{4}}}{\left(\pm(1-y_0\right))^{\frac{1}{4}}}+6 \sqrt{\pi }\sigma_\pm  \frac{\Gamma \left(\frac{3}{4}\right)}{\Gamma \left(\frac{1}{4}\right)}\right]^{-1}\, \nonumber\ .
\end{eqnarray}
Again, the sign depends on whether the theory belongs to $\B_8^+$ or $\B_8^-$. Also $\sigma_+ = 1$ and $\sigma_- = 0$ and
\begin{equation}
h_{\text{IR}} = \frac{64}{w_\pm^3(y_0)}u_s^2\ \mathcal{H}_{\text{IR}} \ .
\end{equation}
Here $\mathcal{H}_{\text{IR}}$ is the parameter appearing in \cite{Faedo:2017fbv}.

\section{Computation of the entanglement entropy of the strip}\label{ap:strip}

In this appendix we give more details on the computation of the entanglement entropy of the strip. Let us first discuss the connected configuration, whose embedding \eqref{eq:embedding_strip1} leads to expression \eqref{eq:EE_strip} which we collect here for ease of reference
\begin{equation}
S_\cup(l) = \frac{V_6 L_y}{4 G_{10}} \int_{-\frac{l}{2}}^{\frac{l}{2}}\dd \sigma^1\  \Xi^{\frac{1}{2}} (1+h\ \dot r ^2)^{\frac{1}{2}}\ ,
\end{equation}
where  $\Xi = h^2 e^{8f+4g-4\Phi}$. There is a conserved quantity in this integral, which can be used to find a simple expression for the embedding:
\begin{equation}\label{eq:FO_embedding}
\dot r = \pm \ h^{-\frac{1}{2}} \sqrt{\frac{\Xi\ }{\Xi_*}-1}\ ,
\end{equation}
where $\Xi_* = \Xi(r_*)$ and the dot indicates differentiation with respect to $\sigma^1$. This allows us to write \eqref{eq:EE_strip} as 
\begin{equation}\label{eq:Scupdiv}
 S_\cup = 2\frac{V_6\ L_y}{4G_{10}} \int_{r_*}^\infty \frac{\Xi\ h^{\frac{1}{2}}}{\sqrt{\Xi-\Xi_*}}\dd r \ .
\end{equation}
As alluded to in Sec.~\ref{sec:strip}, this quantity is UV divergent. Due homogeneity, (\ref{eq:Scupdiv}) possesses the same divergence as the $\sqcup$ configuration \eqref{eq:EE_disconnected}. Then, the difference between them, defined in \eqref{eq:EE_strip_reg} as $\Delta S$, can be computed by performing the integral
\begin{equation}\label{eq:formula_EE_reg}
\begin{aligned}
\Delta S &= S_\cup -S_\sqcup =\frac{V_6\ L_y}{4G_{10}}\left[\ 2\int_{r_*}^\infty \left[\frac{\Xi^{\frac{1}{2}}}{\sqrt{\Xi-\Xi_*}}-1\right] \ 
\Xi^{\frac{1}{2}}h^{\frac{1}{2}}\ \dd r \ -\ 2\int_{r_s}^{r_*}\ \Xi^{\frac{1}{2}}h^{\frac{1}{2}}\ \dd r\ \right] \ .
\end{aligned}
\end{equation}
Interestingly, \eqref{eq:FO_embedding} also allows us to write the width of the strip as
\begin{equation}\label{eq:widthStrip}\begin{aligned}
l &= \int_{-\frac{l}{2}}^{\frac{l}{2}}\ \dd x^1\ = \ 2\int_{r_*}^\infty\frac{\Xi_*^{\frac{1}{2}}}{\sqrt{\Xi-\Xi_*}} h^{\frac{1}{2}}\ \dd r\ ,
\end{aligned}
\end{equation}
in such a way that scanning the parameter space of the turning point of the embedding $r_*\in (r_s,\infty)$, we find the corresponding values of the entanglement entropy and the strip width by simple integration of \eqref{eq:formula_EE_reg} and \eqref{eq:widthStrip}, respectively. 

Again, for non-vanishing CS level, changing to the coordinate $u$ \eqref{ucoord} and performing the rescalings \eqref{eq:dimlessfunctions}, the charges can be factored out. This allows us to redefine $\Delta S$ in such a way that it does not depend on the charges (or the rank of the gauge groups):
\begin{equation}
\label{eq:dimensionlessEEstrip}
\Delta \overline S \ =\  \frac{4G_{10}}{|Q_k|(4 q_c^2 + 3Q_c|Q_k|)}  \times \frac{\Delta S}{L_yV_6} \ =\  \frac{2^8 \pi^4}{9 \lambda}\cdot \frac{N}{|k|\left(\bar{M}^2+2|k|N_c\right)}
\times \frac{\Delta S}{L_yV_6} \ .
\end{equation}     
We can also define a dimensionless strip width \eqref{eq:widthStrip}, namely
\begin{equation}
\overline l\ =\  \frac{|Q_k|^2}{(4q_c^2+3Q_c|Q_k|)^{\frac{1}{2}}}\cdot l\ = \  \frac{\lambda}{6 \pi  N}\frac{|k|^2}{(\bar M + 2|k| N)^\frac{1}{2}}\cdot l\ .
\end{equation}
Similar expressions can be written in order to factor charges out when CS level is vanishing.

\section{Computation of the entanglement entropy of the disk}\label{ap:disk}

Let us now discuss the computation of the entanglement entropy of disks. As in Appendices \ref{ap:solution} and \ref{ap:strip}, we will be discussing the case when $|Q_k|\neq0$ (\textit{i.e.}~the whole family of $\B_8$ excluding $\Bconf$). An analogous analysis can be performed in the case of vanishing CS level. Because the procedure is conceptually identical, we will not discuss it here, the main difference being that the coordinate $u$ is not well defined when $|Q_k|=0$ and a distinct radial coordinate has to be used in that case.
 
First of all, it is useful to change to $r$ as the integration variable in \eqref{eq:EE_disk}. Doing so, the entanglement entropy of the disk reads
\begin{equation}\label{eq:EE1Ddisk}
S_{\text{disk}} =
\frac{V_6}{4 G_{10}} \ 2\pi \int_{r_*}^{\Lambda|Q_k|}\ \dd r \ (\rho'^2+h )^\frac{1}{2} \ \rho \ \Xi^{\frac{1}{2}} \ .
\end{equation}
Note that, because \eqref{eq:EE1Ddisk} is UV divergent, we have explicitly introduced the cut-off $\Lambda$, which after the regularization will be taken to infinity. The embedding is now given by the function $\rho(r)$, which satisfies the second order differential equation coming from \eqref{eq:EulerLagrange}
\begin{equation}\label{eq:EOMembeddingDisk}
\frac{\dd}{\dd r}\left[\frac{\rho'\ \rho\  \Xi^{\frac{1}{2}}}{\sqrt{\rho'^2 + h}}\right] \ - \ \Xi^{\frac{1}{2}}\ \sqrt{\rho'^2 + h}\ =\ 0\ .
\end{equation}
As we mentioned in Sec.~\ref{sec:disk}, there are three types of solutions to \eqref{eq:EOMembeddingDisk} whose boundary conditions will be discussed below. To simplify notation, it is convenient to define the rescaled quantities
\begin{eqnarray}
\rhot &=& \frac{|Q_k|^2}{(4q_c^2+3Q_c |Q_k|)^{\frac{1}{2}}} \ \rho \nonumber\\
\overline{S}_{\text{disk}} &=& \frac{4G_{10}\ |Q_k|}{(4q_c^2 + 3Q_c|Q_k|)^{\frac{3}{2}}}\times \frac{{S}_{\text{disk}}}{V_6} \ =\  \frac{2^7\pi^3\ |k|}{3^3\ (\bar M + 2|k| N)^\frac{3}{2}} \times \frac{{S}_{\text{disk}}}{V_6} \ .
\end{eqnarray}
Let us first explain how to solve the embedding equation \eqref{eq:EOMembeddingDisk} and then explain the regularization of \eqref{eq:EE1Ddisk}. We use the $u$ coordinate as defined in \eqref{ucoord}. 
We can first consider the UV expansion of the metric functions (see appendix \ref{ap:solution}) and solve \eqref{eq:EOMembeddingDisk} perturbatively about the UV. Our solutions eventually deviate from the D2-brane metric and hence from the expansion in \cite{vanNiekerk:2011yi}, in particular by logarithmic terms. We obtain:
\begin{equation}\label{eq:UVrhot}
\begin{aligned}
\rhot &= c_0 - \frac{8(1-b_0^2)}{45 c_0}u^3 +\frac{16(-2+7b_0^2)}{45c_0}u^4 + \parent{c_5 -\frac{64(-1+21b_0^2)}{315c_0}\log u}u^5 \ +  \\
&+ \parent{\frac{20 c_5}{3}-\frac{8
		\left(1-b_0^2\right){}^2}{2025 c_0^3}-\frac{128 \left(861 b_0^2-181\right)}{2835
		c_0}  -\frac{256 \left(21 b_0^2-1\right) }{189 c_0} \log u } u^6\ +\ldots \ .
\end{aligned}
\end{equation}

Note there are two undetermined parameters, $c_0$ and $c_5$. Taking \eqref{eq:radiusat infty} into consideration, we realize that $c_0$ is the (rescaled) value of the radius of the disk, $\overline R = c_0$. This UV expansion is valid for the three cases represented in Fig.~\ref{fig:disk_conf}, the difference between them being determined by the other boundary condition, as explained in Sec.~\ref{sec:disk}, elaborated upon here:
\begin{itemize}
	\item The embedding from Fig.~\ref{fig:disk_conf}(a) has to satisfy the boundary condition \eqref{eq:disk_BC1}, which in the $u$ coordinate reads
	\begin{equation}
		u(0) = u_* \qquad , \qquad \dot u(0) = 0 .
	\end{equation}
	To this end, we solve $\rhot(u)$ about $u=u_*$ by using a series expansion
	\begin{equation}\label{eq:bounday_condition_disk1}
	\rhot(u)\ =\ (u-u_*)^{\frac{1}{2}} \sum_{k=0}^{\infty} B_k (u-u_*)^k\ .
	\end{equation}
	For a given theory of the $\B_8$ family (\textit{i.e.} for a given $b_0$), all the $B_k$ coefficients are determined in terms of $u_*$. For each choice of $u_*$ we use a shooting technique in order to determine the corresponding value of the UV parameters $c_0$ and $c_5$. More precisely, we impose the expansion \eqref{eq:bounday_condition_disk1} near $u_*$ and integrate up to some value $\epsilon_{\text{UV}}$ where we can trust the UV expansion \eqref{eq:UVrhot} within our numerical precision. Then, using a Newton-Raphson routine, we fixed the values of $c_0$ and $c_5$ which render $\rhot(u)$ continuous and differentiable at $\epsilon_{\text{UV}}$. This procedure should then be repeated for each $u_*$.
	
	Since $c_0$ is essentially the dimensionless radius $\overline R$, scanning values for $u_*\in (0,u_s)$ we get all the embeddings for the RT surfaces associated to disks of radius $\overline R\in (0,\overline R_c)$.
	\item Similarly, embeddings represented in Fig.~\ref{fig:disk_conf}(b) satisfy the analogous boundary condition \eqref{eq:disk_BC1}, which after changing to the $u$ coordinate gives
	\begin{equation}
	\lim_{\rhot\to \rhot_*} u(\rhot)= u_s \ .
	\end{equation}
	We then need to solve the equation for $\rhot(u)$ about $u_s$, which leads to a series expansion of the form
	\begin{equation}
	\rhot(u) = \rhot_* + \sum_{k=1}^\infty C_k (u-u_s)^k \ .
	\end{equation}
	For a given theory, the only free parameter in this expansion is $\rhot_*$. Consequently, after imposing this expansion in the equation for $\rhot(u)$ near $u=u_s$, we can find the corresponding values of $c_0$ (\textit{i.e.} $\overline{R}$) and $c_5$ imposed by a shooting procedure analogous to the aforementioned one. In this case, scanning over all the values for $\overline \rho_*\in (0,\infty)$ leads to the corresponding embeddings of RT surfaces of disks with radius $\overline{R}\in \left(\overline{R}_c,\infty\right)$. Note that this type of embedding is not realized when  $b_0=0$, which is the case of the theory which flows to an IR fixed point.
	\item Finally, there is one further type of embedding we are interested in, pictorially represented in Fig.~\ref{fig:disk_conf}(c). The boundary condition in this case is \eqref{eq:embeddding_teo_disks}, which in the $u$ coordinate is given by
	\begin{equation}
	u(\rhot_*) = u_*\qquad , \qquad \dot u(\rhot_*) = 0
	\end{equation}
	with $u_*\neq u_s$ and $\rhot_*\neq 0$. In this case, for the solution about $u=u_*$ we get
	\begin{equation}
	\rhot(u) = \rhot_* + \sum_{k=1}^\infty\ D_k^{\pm} \ (u-u_*)^{\frac{k}{2}} \ .
	\end{equation}
	As the superscript in $D_k^\pm$ suggests, there are two different series, depending on a choice of sign that has to be made while solving the first coefficient. The rest of the coefficients are fixed in each case as functions of $\rhot_*$ and $u_*$. Each choice is giving a distinct branch of the embedding corresponding to Fig.~\ref{fig:disk_conf}(c). For each of the two branches, the shooting method gives a different value of $c_0$, corresponding to the two radius of the two disks to which this embedding is attached. We refer to them as $\overline{R}_1$ and $\overline{R}_2$. Scanning the parameters space $u_*\in (0,u_s)$ and $\rhot_* \in (0,\infty)$ we efficiently get the embeddings corresponding to all possible values of $\overline R_1$ and $\overline R_2$.
\end{itemize}

Now that we understand the different embeddings we encounter, we can turn to the issue of regulating the action functional. Knowing the UV expansion of all the functions which are involved in the computation, it is possible to study the divergence structure of the entanglement entropy in this particular problem. First, let us write the integral \eqref{eq:EE1Ddisk} in the $u$ coordinate:
\begin{equation}\label{eq:actionDisk}
\overline S_{\text{disk}} =  \int_{u_*}^{\Lambda^{-1}} \dd u\parent{-\frac{1}{u^2}\ \Xi^{\frac{1}{2}}\ \rhot\ (\mathbf{h}+u^4\rhot'^2)^\frac{1}{2} } \equiv \int_{u_*}^{\Lambda^{-1}} \dd u\  L_D \ ,
\end{equation}
where $u_* = |Q_k| \ r_*^{-1}$  is the value of the radial coordinate at the turning point (or $u_*=u_s$ if the embedding reaches the bottom of the geometry) and $\overline{\Xi}$ is the dimensionless version of $\Xi$, namely 
\begin{equation}
\overline \Xi \ =\ \frac{4q_c^2 + 3Q_c |Q_k|}{|Q_k|^6} \ \mathbf{h}^2 e^{8\FF+4\GG-4\Phi}\ .
\end{equation}
Also, last equality in \eqref{eq:actionDisk} defines $L_D$. Replacing the functions by their UV expression and performing the integral we get
\begin{equation}
\begin{aligned} \label{eq:actionUVexpandsion}
\overline S_{\text{disk}} =& \ \int_{u_*}^{\Lambda^{-1}} \dd u\parent{-\frac{1}{u^2}\ \overline \Xi^{\frac{1}{2}}\ \rhot\ (\mathbf{h}+u^4\rhot'^2)^\frac{1}{2} } \\[2mm]
 =&\  c_0\, \Big[\, \frac{1}{30} \left(1-b_0^2\right) \Lambda ^2 \ - \ \frac{4}{45} \left(4
b_0^2+1\right) \Lambda\  - \  \frac{4}{63} \left(21 b_0^2-1\right)
\log \Lambda \ + \ S_0  \\[2mm]
&\ +\ \frac{2 \left(14 b_0^2 \left(720
	c_0^2-1\right)+7 b_0^4-480 c_0^2+7\right)}{4725 c_0^2 
} \frac{1}{\Lambda}\ +\ \OO(\Lambda^{-2})\ \Big].
\end{aligned}
\end{equation}

A few important remarks are in order:
\begin{itemize}
	\item The leading divergence in \eqref{eq:actionUVexpandsion} is of the same order as the one found in \cite{vanNiekerk:2011yi}, namely $\Lambda^2$.
	\item Because we deviate from D2-brane asymptotics eventually \eqref{eq:solutionUV}, there are new terms appearing. Nevertheless, the counterterms are readily available
	\begin{equation}\label{eq:counterterm}
	\overline  S_{\text{disk}}^{\text{ct}} (\Lambda^{-1})\ =\ c_0 \, \Big[\,  \frac{1}{30} \left(1-b_0^2\right) \Lambda ^2\ -\ \frac{4}{45} \left(4
	b_0^2+1\right) \Lambda \ -\ \frac{4}{63} \left(21 b_0^2-1\right)
	\log \Lambda \, \Big].
	\end{equation}
	\item In \eqref{eq:actionUVexpandsion} we see that the finite contribution $S_0$ is just the integration constant and we can set it to zero (or absorb it in \eqref{eq:counterterm}). One can show that this corresponds to picking a particular renormalization scheme on the field theory \cite{Hoyos:2016cob,Ecker:2017fyh}.
	\item Because there is a global $c_0$ multiplying \eqref{eq:counterterm}, this tells us that at the end of the day $S_{\text{disk}}^{\text{ct}}$ is proportional to the perimeter of the disk, thus fulfilling and area law.
	\item Note also that the counterterms do not depend on the precise embedding. The counterterms depend on $c_0$, which is essentially the radius of the disk, but not on $c_5$, since its value is determined after solving the equation of the embedding and therefore depend on the IR data.
	\item Finally, note that the counterterms not only are independent of $c_5$, but also all the subleading parameters $f_4$ and $b_4$ appearing in \eqref{eq:solutionUV}.
	This means that the counterterms are indeed completely independent of the subleading parameters of the UV expansions, which are associated to VEVs and determined by imposing some IR condition.
\end{itemize}

Taking into account that we can express the counterterms in \eqref{eq:counterterm} as
\begin{equation}
\label{eq:regDiskFromLagrangian}
\begin{aligned}
\overline  S_{\text{disk}}^{\text{ct}} (\Lambda^{-1}) = \int_{u_*}^{\Lambda^{-1}}\dd u \ &L_D^{\text{ct}}  \ +\  \overline  S_{\text{disk}}^{\text{ct}} (u_*)\\
\mbox{where }\qquad &L_D^{\text{ct}}  \equiv \frac{\left(b_0^2-1\right) c_0}{15 u^3}+\frac{4 \left(4 b_0^2+1\right) c_0}{45
	u^2}+\frac{4 \left(21 b_0^2-1\right) c_0}{63 u}\ ,
\end{aligned}\end{equation}
the regularized entanglement entropy of the disk can be written in a way that makes the numerical computations easier. This yields
\begin{equation}\label{eq:formulaEE}
\overline  S_{\text{disk}}^{\text{reg}} = \lim_{\Lambda\to\infty} \left[\overline S_{\text{disk}} - \overline S_{\text{disk}}^{\text{ct}}(\Lambda^{-1})\right] = \int_{u_*}^{0} \dd u \parent{L_D-L_D^{\text{ct}}} -S_{\text{disk}}^{\text{ct}}(u_*)\ .
\end{equation}
Finally, we use \eqref{eq:formulaEE} in order to define a version of $\ffunc_{\text{disk}}$ where the charges has been factored out
\begin{equation}
\label{eq:Ffunction_disk_dimless}
\overline\ffunc_{\text{disk}}(\overline R) =\overline R\cdot \frac{\dd \overline S^{\text{reg}}_{\text{disk}}}{\dd \overline R} - \overline S^{\text{reg}}_{\text{disk}} \ .
\end{equation}

\end{document}